\documentclass[twocolumn,english,aps,pra,showpacs,superscriptaddress]{revtex4-2}
\usepackage{color}
\usepackage{verbatim}
\usepackage{textcomp}
\usepackage{mathrsfs}
\usepackage{amsmath}
\usepackage{amssymb}
\usepackage{graphicx}
\usepackage{wasysym}
\usepackage{esint}

\makeatletter

\providecommand{\tabularnewline}{\\}

\makeatother

\usepackage{babel}
\begin{document}
\title{An efficient variational polaron master equation for non-Markovian spin-boson dynamics: Transformed initial states and general observables}
\author{Chien-Chang Chen}
\affiliation{Department of Physics and Center for Theoretical Physics, National
Taiwan University, Taipei 106319, Taiwan}
\affiliation{Center for Quantum Science and Engineering, National Taiwan University,
Taipei 106319, Taiwan}
\affiliation{Physics Division, National Center for Theoretical Sciences, Taipei
106319, Taiwan}
\author{Hsi-Sheng Goan}
\email{goan@phys.ntu.edu.tw}

\affiliation{Department of Physics and Center for Theoretical Physics, National
Taiwan University, Taipei 106319, Taiwan}
\affiliation{Center for Quantum Science and Engineering, National Taiwan University,
Taipei 106319, Taiwan}
\affiliation{Physics Division, National Center for Theoretical Sciences, Taipei
106319, Taiwan}
\date{\today}
\begin{abstract}
We investigate the effects of initial states on the dynamics of a
quantum spin-boson model using an efficient variational polaron master
equation approach. The transformation of a factorized system-environment
initial state into the variational polaron frame naturally introduces
inhomogeneous terms in the master equation, which are essential for
accurately describing non-Markovian dynamics beyond the weak system-environment
coupling regime. To efficiently account for these effects, we formulate
the dynamics in an extended auxiliary Liouville space, converting
the resulting integro-differential equations into coupled ordinary
differential equations that can be solved efficiently. We further
introduce a simplified variational polaron technique for computing
the time evolution of expectation values of general observables, including
those that do not commute with the polaron transformation, thereby
overcoming a major limitation of previous approaches and enabling
the unified treatment of both coherence and population dynamics. The
effects of transformed initial states generated from arbitrary factorized
system-environment initial states are systematically characterized
through expectation values and trace-distance analysis, providing
a comprehensive picture of the associated bath-relaxation dynamics
over a broad range of qubit tunneling amplitudes, temperatures, and
system-bath coupling strengths. Comparison with the weak-coupling
master equation reveals a delay phenomenon associated with multi-bath-quanta
processes, which becomes pronounced at intermediate coupling strengths
and tunneling amplitudes. We further show that coherent superposition
initial states induce stronger nonequilibrium effects than localized
states, with the inhomogeneous terms playing a crucial role in the
short-time dynamics. The proposed formalism is applicable to both
undriven and resonantly driven spin-boson models, providing a computationally
efficient framework for investigating open quantum-system dynamics
with arbitrary factorized initial states and general physical observables
across a wide parameter regime.
\end{abstract}
\maketitle

\section{Introduction}

The dynamics of open quantum systems---quantum systems interacting
with environmental degrees of freedom---plays a central role in modern
quantum science. Understanding how environmental interactions influence
the reduced dynamics of a quantum system is essential for a broad
range of applications, including quantum information processing, quantum
control, quantum thermodynamics, and quantum transport. Of particular
interest is the accurate description of both population dynamics,
associated with the diagonal elements of the reduced density matrix,
and coherence dynamics, associated with the off-diagonal elements.
These quantities determine the evolution of quantum states and are
directly related to the experimentally observable properties of quantum
systems.

A widely used approach to describe open quantum dynamics is the
quantum master equation derived from second-order perturbation theory,
in which the system-environment interaction is treated as a weak perturbation
\citep{Scully97,Carmichael,Gardiner00,Milburn08,Breuer02}. Depending
on the approximations employed, these master equations can be formulated
within either Markovian frameworks, which neglect environmental memory
effects \cite{Scully97,Carmichael,Gardiner00,Milburn08,Goan01,*Goan01_2,*Goan01_3,*Goan01_4,Utami04,*Utami04_3,*Utami04_2,Hu15},
or non-Markovian frameworks that retain memory effects arising from
the environment \cite{Breuer02,Liu07,Ali10,Chen11,Goan10,*Goan10_2,Optimal_control,Chou15,initial_paper}.
Such approaches have achieved considerable success in the weak-coupling
regime and constitute the foundation of many theoretical descriptions
in quantum optics and open quantum systems.

However, weak-coupling master equations become inadequate when the
system-environment interaction is sufficiently strong or when thermal
fluctuations are comparable to the characteristic energy scales of
the system. In these regimes, higher-order processes involving multiple
environmental quanta become important and cannot be accurately captured
within a second-order perturbative treatment. Although numerically exact approaches,
such as the quasiadiabatic propagator path integral (QUAPI) \citep{A_general_RWA_VP_master_equation,Accuracy_of_variational_polaron_eq,Criteria_for_small_polarion}
and the hierarchical equations of motion (HEOM) \citep{hierarchal_EOM_initial,HEOM_Huang2023},
provide reliable descriptions beyond the perturbative regime,
their computational cost often limits their applicability and obscures
the underlying physical mechanisms.

An alternative approach is provided by the polaron transformation
\citep{Criteria_for_small_polarion,SP-called_McCutcheonNazir2011,SP-called_Jang2011,SP-called_ChengSilbey2008,SP-called_Holstein1959,Full_polaron_steady_state_2020},
which incorporates a substantial part of the system-environment interaction
into the unperturbed Hamiltonian through a unitary transformation.
By dressing the system with coherent displacements of the environmental
modes, the polaron framework naturally extends the validity of perturbative
master equations into regimes of stronger coupling and higher temperatures.
The full-polaron master equation has successfully explained experimental
observations in systems such as GaAs double quantum dots \citep{step-like-experiment2014,Full_polaron_steady_state_2020},
demonstrating that realistic devices may operate beyond the conventional
weak-coupling regime. Furthermore, introducing an optimizable variational
parameter into the transformation leads to the variational polaron
master equation \citep{VP_Master_Eq_DC_and_Imhomo,A_general_RWA_VP_master_equation},
which smoothly interpolates between the weak-coupling and full-polaron
limits, thus providing a unified description across a broad range
of system-environment coupling strengths. However, some important challenges
remain nontrivial within the variational polaron approach. In this
work, we address these challenges using the spin-boson model, which
describes a two-level system interacting with a bosonic environment
and is relevant to a variety of physical systems, including semiconductor
quantum dots \citep{Full_polaron_steady_state_2020}, molecular systems
\citep{Criteria_for_small_polarion}, and superconducting qubits \citep{Superconducting_Qubit_Magazzu2018}.

Within the variational polaron framework, the population dynamics
characterized by the expectation value $\left\langle \sigma_{z}\left(t\right)\right\rangle $
can be obtained directly from the reduced density matrix in the transformed
frame. This follows from the fact that the variational polaron transformation
commutes with $\sigma_{z}$, leaving the operator invariant. In contrast,
the coherence observables $\sigma_{x}$ and $\sigma_{y}$ do not commute
with the transformation and are consequently mapped onto system-bath
operators. Their expectation values therefore involve additional bath
correlation functions and cannot be evaluated solely from the reduced
density matrix in the transformed frame. As a result, obtaining the
 coherence dynamics of the laboratory-frame, $\left\langle \sigma_{x}\left(t\right)\right\rangle $
and $\left\langle \sigma_{y}\left(t\right)\right\rangle $, remains
considerably more challenging than calculating population dynamics.
This limitation has restricted the application of polaron-based master
equations to problems that involve transverse spin measurements \citep{superposition_state_2008,superposition_state1_2009},
optical coherence, and resonance fluorescence, including the Mollow
triplet \citep{variational_polaron_RWA_two_time_correlation}.

An equally important challenge concerns the treatment of the initial
state. In the laboratory frame, it is commonly assumed that the total density operator is
 factorized as $\rho_{T}\left(0\right)=\rho_{s}(0)\otimes\rho_{b}$,
where the environment is initially in thermal equilibrium. However, after the
variational polaron transformation, this factorized state
becomes correlated through bath displacements, giving rise to inhomogeneous
terms in the master equation that describe the relaxation of the displaced
bath toward equilibrium\citep{Criteria_for_small_polarion,VP_Master_Eq_DC_and_Imhomo,coherent_initial_state_inhomo_SP}.
When the system is initialized in a coherent superposition of its
basis states, transformed off-diagonal density-matrix elements
introduce an additional type of bath displacement, further increasing
the complexity of inhomogeneous contributions \citep{coherent_initial_state_inhomo_SP}.
Consequently, incorporating general transformed initial states into
the variational polaron master equation presents a significant theoretical
challenge.

Previous studies have addressed these issues within the full-polaron
framework by developing a general treatment of transformed observables
and examining the influence of transformed initial states on excitonic
dynamics in multichromophoric systems \citep{coherent_initial_state_inhomo_SP,multisite_SP_time-local_initial_fac_in_imhomo,off_diagonal_expectaion_sp_2012,incoherent_initial_state_inhomo_correlated_FP_2019,coherent_correlated_initial_state_inhomo_FP_2019}.
These studies demonstrated that nonequilibrium bath relaxation can
significantly modify the system dynamics and that coherent initial
states generally produce stronger effects than localized ones. More
recently, initially correlated states prepared from the thermal equilibrium
state of the total system have been investigated using a variational
polaron master equation \citep{Qin2025}; however, that study focused
on population dynamics and did not address the dynamics of transformed coherence observables.
Inhomogeneous terms induced by the initial bath state and corrections to transformed
coherence observables have also been incorporated into the variational polaron framework, with explicit applications to the quantum regression theorem and single-time dynamics, but only for an initial diagonal system state \citep{BundgaardNielsen2026}.
A general variational polaron treatment of factorized laboratory-frame system-environment 
initial states with off-diagonal system components, together with
an efficient formulation for consistently evaluating the resulting
transformed dynamics and observables, remains to be developed.

In this work, we develop a unified variational polaron framework that
simultaneously incorporates transformed initial states and transformed
observables. We extend the variational polaron master equation by
including the complete polaron-transformed initial state (with both diagonal and off-diagonal components of the system state) together
with the associated inhomogeneous terms arising from bath relaxation
out of displaced nonequilibrium configurations \citep{VP_Master_Eq_DC_and_Imhomo,coherent_initial_state_inhomo_SP}.
At the same time, we establish a systematic approach for evaluating
expectation values of observables that do not commute with the variational
polaron transformation, thereby providing direct access to both population
and coherence dynamics in the laboratory frame.

To overcome the computational complexity associated with the resulting
integro-differential equations, we formulate the dynamics in an extended
auxiliary Liouville space. By expressing the bath correlation functions
as a sum of exponentials \citep{Optimal_control,Optimal_control2,initial_paper},
 nonlocal memory kernels are transformed into a set of coupled
local-in-time differential equations involving auxiliary matrices.
This reformulation significantly simplifies the numerical implementation
while naturally accommodating both homogeneous and inhomogeneous contributions
within a unified framework.

Using this formalism, we investigate the influence of transformed
initial states on the dynamics of the spin-boson model through the
evolution of $\left\langle \sigma_{z}\left(t\right)\right\rangle $,
$\left\langle \sigma_{x}\left(t\right)\right\rangle $ and $\left\langle \sigma_{y}\left(t\right)\right\rangle $,
together with trace-distance analysis that quantifies the effects
of bath relaxation. We further compare the variational polaron and
weak-coupling master equations over a broad range of system parameters,
clarifying the role of higher-order system-environment interactions
beyond second-order perturbation theory. Finally, we demonstrate that
the formalism can be directly extended to a resonantly driven spin-boson
model, enabling the investigation of decoherence dynamics in both
 laboratory and rotating frames \citep{NV_center_measure_in_rotatng_frame}.
Our results for representative cases of both the undriven and resonantly
driven spin-boson models are further benchmarked against numerically exact HEOM calculations as implemented in Ref. \citep{HEOM_Huang2023} to confirm their validity.

The remainder of this paper is organized as follows. In Sec. \ref{sec:spin_boson_model},
we introduce the spin-boson model and derive the variational polaron
master equation together with the transformed initial state. Section
\ref{sec:expectation_value} presents the formalism for evaluating
transformed observables in the laboratory frame. In Sec. \ref{sec:Initial_state_trace_distance},
we discuss the approximations of the transformed initial state and analyze
their validity using the trace distance. Section \ref{sec:Dynamics_result_undriven}
presents numerical results for the undriven spin-boson model, while
Sec. \ref{sec:driven_spin_boson} extends the analysis to the resonantly
driven case. Section \ref{sec:Conclusion} provides the concluding
remarks. Additional details on the bath correlation functions and
auxiliary matrices, the pure-dephasing limit, and the HEOM benchmark
are provided in the Appendices.

\section{Model Hamiltonian and non-Markovian master equation}

\label{sec:spin_boson_model}

\subsection{Spin-boson model}

We consider the spin-boson model, which describes a two-level system
coupled to a bosonic environment. The total Hamiltonian is given by
$\left(\hbar=1\right)$: 
\begin{eqnarray}
H_{T} & = & H_{s}+H_{b}+H_{sb},\label{H_T}\\
H_{s} & = & \frac{\varepsilon}{2}\sigma_{z}+\frac{\Delta}{2}\sigma_{x},\label{H_s}\\
H_{b} & = & \sum_{k}\omega_{k}b_{k}^{\dagger}b_{k},\\
H_{sb} & = & \sigma_{z}\sum_{k}g_{k}\left(b_{k}^{\dagger}+b_{k}\right).\label{eq:H_sb}
\end{eqnarray}
Here, $\varepsilon$ is the energy splitting between the excited state
$\left|1\right\rangle $ and the ground state $\left|0\right\rangle $,
and $\Delta$ denotes the tunneling amplitude between the two states.
The Pauli operators are defined by $\sigma_{z}=\left|1\right\rangle \left\langle 1\right|-\left|0\right\rangle \left\langle 0\right|$
and $\sigma_{x}=\left|0\right\rangle \left\langle 1\right|+\left|1\right\rangle \left\langle 0\right|$.
We also introduce the Pauli operator $\sigma_{y}=i\left(\left|0\right\rangle \left\langle 1\right|-\left|1\right\rangle \left\langle 0\right|\right)$,
which will be used in subsequent sections. The bath Hamiltonian $H_{b}$
describes a collection of harmonic oscillators with frequencies $\omega_{k}$,
where $b_{k}^{\dagger}\left(b_{k}\right)$ is the corresponding creation
(annihilation) operator. The interaction Hamiltonian $H_{sb}$ represents
the coupling between the two-level system and the bath, with coupling
strength $g_{k}$ for the $k$th bath mode.

\subsection{Weak-coupling master equation}

\label{sec:weak-coupling}

We derive a time-local master equation that is accurate to second
order in the system-bath interaction using the time-convolutionless
(TCL) projection-operator formalism \citep{Breuer02,time-local_vs_non-loca_2nd_ME_initial}.
When the tunneling term is absent ($\Delta=0$), the Hamiltonian in
Eqs. (\ref{H_T})-(\ref{eq:H_sb}) reduces to the pure-dephasing spin-boson
model, whose dynamics is exactly described by a time-local (time-convolutionless)
master equation. Motivated by this property, we employ the TCL projection-operator
formalism to describe the dynamics of the reduced density matrix $\rho_{s}\left(t\right)=\textrm{tr}_{b}\left[\rho_{T}\left(t\right)\right]$
for the present model.

The projection operator $P$ is defined by
\[
P\rho_{T}(t)=\textrm{tr}_{b}[\rho_{T}(t)]\otimes\rho_{b}=\rho_{s}(t)\otimes\rho_{b}
\]
where $\rho_{b}=e^{-\beta H_{b}}/\textrm{tr}_{b}e^{-\beta H_{b}}$
is the thermal equilibrium state of the bath at temperature $k_{B}T=1/\beta$.
The complementary projection operator is defined by $Q=\mathcal{I}-P$,
where $\mathcal{I}$ denotes the identity superoperator in the total
system-bath Hilbert space.

Applying the projection operators to the interaction-picture Liouville
equation,
\begin{equation}
\dot{\tilde{\rho}}_{T}\left(t\right)=-i\left[\tilde{H}_{sb}\left(t\right),\tilde{\rho}_{T}\left(t\right)\right],\label{eq:interaction_Liouville_eq}
\end{equation}
and tracing over the bath degrees of freedom yield the equation of
motion for the reduced interaction-picture density matrix, $\tilde{\rho}_{s}\left(t\right)=\textrm{tr}_{b}\left[\tilde{\rho}_{T}\left(t\right)\right]$,
as
\begin{eqnarray}
\frac{\partial}{\partial t}\tilde{\rho}_{s}\left(t\right) & = & -i\textrm{tr}_{b}\left[\tilde{H}_{sb}\left(t\right),\mathcal{I}\tilde{\rho}_{T}\left(t\right)\right]\label{dt_rho_s(t)_inter_pic_I}\\
 & = & \textrm{tr}_{b}\left[\tilde{\mathscr{L}}_{sb}\left(t\right)Q\tilde{\rho}_{T}\left(t\right)\right].\label{dt_rho_s(t)_inter_pic_Q}
\end{eqnarray}
Here, we have inserted an identity operator $\mathcal{I}=P+Q$ in
Eq.~(\ref{dt_rho_s(t)_inter_pic_I}). The interaction-picture Liouvillian
is defined by $\tilde{\mathscr{L}}_{sb}\left(t\right)A\equiv-i\left[\tilde{H}_{sb}\left(t\right),A\right]$,
where $\tilde{H}_{sb}\left(t\right)=\mathcal{G}_{s}\left(0,t\right)\sigma_{z}B\left(t\right)$,
with $B\left(t\right)=\sum_{k}g_{k}\left(b_{k}^{\dagger}e^{i\omega_{k}t}+b_{k}e^{-i\omega_{k}t}\right)$.
In deriving Eq.~(\ref{dt_rho_s(t)_inter_pic_Q}), we use
\begin{eqnarray*}
\textrm{tr}_{b}\left[B\left(t\right),P\tilde{\rho}_{T}\left(t\right)\right] & = & \textrm{tr}_{b}\left[B\left(t\right),\tilde{\rho}_{s}\left(t\right)\otimes\rho_{b}\right]=0,
\end{eqnarray*}
which follows from the fact that the thermal equilibrium state satisfies
$\textrm{\ensuremath{\textrm{tr}_{b}}}\left(b_{k}\rho_{b}\right)=\textrm{tr}_{b}\left(b_{k}^{\dagger}\rho_{b}\right)=0$
for a thermal bath. The propagator superoperator is defined by
\begin{equation}
\mathcal{G}_{j}\left(t,t'\right)\equiv\textrm{T}_{+}\exp\left[\intop_{t'}^{t}\mathscr{L}_{j}\left(t''\right)dt''\right],\label{eq:G}
\end{equation}
where $\textrm{T}_{+}$ denotes the time-ordering operator for an
explicitly time-dependent Hamiltonian \citep{Breuer02}. The Liouville
superoperator is defined as
\begin{equation}
\mathscr{L}_{j}A\equiv-i\left[H_{j},A\right],\label{eq:L}
\end{equation}
with $j=s$ in the present context.

Expanding $Q\tilde{\rho}_{T}\left(t\right)$ to first order in the
interaction Liouvillian gives
\begin{eqnarray}
Q\tilde{\rho}_{T}\left(t\right) & = & Q\tilde{\rho}_{T}\left(0\right)+\int_{0}^{t}\tilde{\mathscr{L}}_{sb}\left(t'\right)Q\tilde{\rho}_{T}\left(0\right)dt'\nonumber \\
 &  & +\int_{0}^{t}\tilde{\mathscr{L}}_{sb}\left(t'\right)\tilde{\rho}_{s}\left(t\right)\otimes\rho_{b}dt'.\label{Q_rho_T_1st}
\end{eqnarray}
Substituting this expression into Eq. \eqref{dt_rho_s(t)_inter_pic_Q}
yields the second-order TCL master equation in the interaction picture
\citep{Breuer02}:
\begin{eqnarray}
\dot{\tilde{\rho}}_{s}\left(t\right) & = & \textrm{tr}_{b}\left[\tilde{\mathscr{L}}_{sb}\left(t\right)Q\tilde{\rho}_{T}\left(0\right)\right]\nonumber \\
 &  & +\int_{0}^{t}\textrm{tr}_{b}\left[\tilde{\mathscr{L}}_{sb}\left(t\right)\tilde{\mathscr{L}}_{sb}\left(t'\right)Q\tilde{\rho}_{T}\left(0\right)\right]dt'\nonumber \\
 &  & +\int_{0}^{t}\textrm{tr}_{b}\left[\tilde{\mathscr{L}}_{sb}\left(t\right)\tilde{\mathscr{L}}_{sb}\left(t'\right)\tilde{\rho}_{s}\left(t\right)\otimes\rho_{b}\right]dt'.\label{weak master Eq interaction pic}
\end{eqnarray}
We assume an initially factorized state, $\rho_{T}\left(0\right)=\rho_{s}(0)\otimes\rho_{b}$,
where the system may initially be in either a pure or mixed state.
Consequently, $Q\tilde{\rho}_{T}\left(0\right)=0$, and the inhomogeneous
terms in Eq. (\ref{weak master Eq interaction pic}) vanish.

Tracing over the bath degrees of freedom and transforming back to
the Schr\"odinger picture, we obtain the weak-coupling master equation,
\begin{eqnarray}
\dot{\rho}_{s}\left(t\right) & = & \mathscr{L}_{s}\rho_{s}\left(t\right)+\mathscr{L}_{z}\left[\mathcal{D}\left(t\right)\rho_{s}\left(t\right)+\textrm{H.c.}\right],\label{weak coupling master eq}
\end{eqnarray}
where
\begin{eqnarray}
\mathcal{D}\left(t\right) & = & -i\int_{0}^{t}C\left(t-t'\right)\mathcal{G}_{s}(t,t')\sigma_{z}dt'.\label{eq:D(t)}
\end{eqnarray}
Here, H.c. denotes the Hermitian conjugate, and the Liouvillian associated
with the operator ($\sigma_{z}$) is defined as $\mathscr{L}_{z}A=-i\left[\sigma_{z},A\right]$.
The bath correlation function $C\left(\tau\right)$ is given by \citep{Carmichael,Breuer02}
\begin{eqnarray}
C\left(\tau\right) & \equiv & \textrm{tr}_{b}\left[B\left(0\right)B\left(-\tau\right)\rho_{b}\right]\nonumber \\
 & = & \int_{0}^{\infty}d\omega J\left(\omega\right)\left[\cos\left(\omega\tau\right)\coth\left(\frac{\beta\omega}{2}\right)-i\sin\left(\omega\tau\right)\right],\nonumber \\
\label{C(tau)}
\end{eqnarray}
where the spectral density is defined as $J\left(\omega\right)=\sum_{k}\left|g_{k}\right|^{2}\delta\left(\omega-\omega_{k}\right)$.

To solve Eqs.~(\ref{weak coupling master eq}) and (\ref{eq:D(t)})
without introducing further approximations, we expand the bath correlation
function as a sum of exponentials \citep{Optimal_control,Optimal_control2,initial_paper,Full_polaron_steady_state_2020},
\begin{equation}
C\left(\tau\right)=\sum_{m}\alpha_{m}e^{\gamma_{m}\tau},\label{eq:C_m}
\end{equation}
where complex coefficients $\alpha_{m}$ and $\gamma_{m}$ are determined
numerically. Substituting this expansion into Eq.~(\ref{eq:D(t)})
gives $\mathcal{D}\left(t\right)=\sum_{m}\mathcal{D}_{m}\left(t\right)$,
with $\mathcal{D}_{m}\left(t\right)=-i\int_{0}^{t}\alpha_{m}e^{\gamma_{m}\left(t-t'\right)}\mathcal{G}_{s}\left(t,t'\right)\sigma_{z}dt'$.
Using the relation $\partial\mathcal{G}_{s}\left(t,t'\right)/\partial t=\mathscr{L}_{s}\mathcal{G}_{s}\left(t,t'\right)$,
Eqs.~(\ref{weak coupling master eq}) and (\ref{eq:D(t)}) can be
transformed into a set of coupled linear equations,
\begin{eqnarray}
\dot{\rho}_{s}\left(t\right) & = & \mathscr{L}_{s}\rho_{s}\left(t\right)+\mathscr{L}_{z}\sum_{m}[\mathcal{D}_{m}\left(t\right)\rho_{s}(t)+\textrm{H.c.}],\label{eq:wcME}\\
\dot{\mathcal{D}}_{m}\left(t\right) & = & \left[\mathscr{L}_{s}+\gamma_{m}\right]\mathcal{D}_{m}\left(t\right)-i\alpha_{m}\sigma_{z}.\label{eq:Dm}
\end{eqnarray}
Equations (\ref{eq:wcME}) and (\ref{eq:Dm}) constitute the weak-coupling
time-local master equation. Since the Born approximation is the only
approximation employed in the derivation, this master equation is
valid in the weak system-bath coupling regime.

\subsection{Variational polaron transformation}

\label{sec:Variational-polaron_transformation}

To extend the treatment beyond the weak-coupling regime, we perform
the variational polaron transformation \citep{S-H_variational,VP_Master_Eq_DC_and_Imhomo},
\begin{equation}
H_{T}'=e^{V}H_{T}e^{-V},\label{eq:Var_polaron_transform}
\end{equation}
where $V=\Phi\sigma_{z}/2$, with 
\begin{equation}
\Phi=2\sum_{k}\frac{g_{k}}{\omega_{k}}\xi_{k}\left(b_{k}^{\dagger}-b_{k}\right),\label{eq:Phi}
\end{equation}
and $\xi_{k}$ is a variational parameter to be determined later.
Under this transformation, the Hamiltonian becomes
\begin{equation}
H_{T}'=H_{s}'+H_{sb}'+H_{b},\label{H'_T}
\end{equation}
where
\begin{eqnarray}
H_{s}' & = & \frac{\epsilon}{2}\sigma_{z}+\frac{\eta\Delta}{2}\sigma_{x}+R\sigma_{0},\label{H'_s}\\
H_{sb}' & = & \sum_{i}\varLambda_{i}B_{i},\label{H'_sb}
\end{eqnarray}
Here, $\sigma_{0}$ denotes the identity operator, and the transformed
system operators are
\begin{eqnarray}
\varLambda_{1} & = & \frac{\Delta}{2}\sigma_{x},\label{eq:Garmma_1}\\
\varLambda_{2} & = & \frac{\Delta}{2}\sigma_{y},\\
\varLambda_{3} & = & \sigma_{z},
\end{eqnarray}
while the corresponding bath operators are
\begin{eqnarray}
B_{1} & = & \cosh\Phi-\eta,\label{eq:B_1}\\
B_{2} & = & i\sinh\Phi,\label{eq:B_2}\\
B_{3} & = & \sum_{k}g_{k}\left(1-\xi_{k}\right)\left(b_{k}^{\dagger}+b_{k}\right),\label{eq:B_3}
\end{eqnarray}
with the energy shift
\begin{equation}
R=\sum_{k}\frac{g_{k}^{2}}{\omega_{k}}\xi_{k}\left(\xi_{k}-2\right).\label{R(xi_k)}
\end{equation}
influences the determination of $\xi_{k}$.

The variational polaron transformation transfers part of the tunneling
term into the system-bath interaction. In particular, the transformed
tunneling operator contains the factor $\left(\Delta/2\right)\sigma_{x}\cosh\Phi$,
which describes spin tunneling accompanied by collective displacements
of the bath modes. To retain the coherent contribution of this process
in the unperturbed system Hamiltonian, we separate its bath average
and define the renormalized tunneling amplitude through
\begin{eqnarray}
\eta & = & \left\langle \cosh\Phi\right\rangle _{H_{b}}\nonumber \\
 & \equiv & \textrm{tr}_{b}\left(\rho_{b}\cosh\Phi\right)\nonumber \\
 & = & \exp\left[-2\sum_{k}\left(\frac{g_{k}}{\omega_{k}}\xi_{k}\right)^{2}\coth\left(\frac{\beta\omega_{k}}{2}\right)\right].\label{eta(xi_k)}
\end{eqnarray}
Accordingly, the fluctuating part is absorbed into the interaction
Hamiltonian through the definition of $B_{1}$ in Eq. (\ref{eq:B_1}),
while the bath-averaged contribution, $\left(\eta\Delta/2\right)\sigma_{x}$,
is incorporated into the renormalized system Hamiltonian $H_{s}'$.
The factor $\eta$ therefore determines the effective tunneling strength
after dressing by the bosonic environment.

The renormalization factor admits a simple physical interpretation
in terms of the overlap between the spin-dependent displaced bath
states \citep{multipolaron}. The variational ground state is
\begin{eqnarray}
\left|\Psi\right\rangle  & = & e^{-V}\left(\left|G\right\rangle \otimes\left|\textrm{vac}\right\rangle \right)\nonumber \\
 & = & p_{-}\left|1\right\rangle \otimes\left|f_{-}\right\rangle -p_{+}\left|0\right\rangle \otimes\left|f_{+}\right\rangle ,\label{eq:ground_state}
\end{eqnarray}
where $\left|G\right\rangle =p_{-}\left|1\right\rangle -p_{+}\left|0\right\rangle $
is the ground state of the bare system Hamiltonian $H_{s}$ satisfying
$p_{+}^{2}+p_{-}^{2}=1$, $\left|\textrm{vac}\right\rangle $ is the
bosonic vacuum state, and $\left|f_{\pm}\right\rangle =e^{\pm\Phi/2}\left|\textrm{vac}\right\rangle $
are multimode coherent states. Their overlap is
\begin{equation}
\left\langle f_{\mp}\right|\left.f_{\pm}\right\rangle =\left\langle \textrm{vac}\right|e^{\pm\Phi}\left|\textrm{vac}\right\rangle =\exp\left[-2\sum_{k}\left(\frac{g_{k}}{\omega_{k}}\xi_{k}\right)^{2}\right],
\end{equation}
which directly determines the spin coherence, $\left\langle \Psi\right|\sigma_{x}\left|\Psi\right\rangle =-2p_{+}p_{-}\left\langle f_{+}\right|\left.f_{-}\right\rangle $.
At finite temperature, replacing the vacuum projector by the thermal
equilibrium state $\rho_{b}$ gives
\begin{equation}
\textrm{tr}_{b}\left(e^{\pm\frac{\Phi}{2}}\rho_{b}e^{\pm\frac{\Phi}{2}}\right)=\left\langle e^{\pm\Phi}\right\rangle _{H_{b}}=\eta,\label{eq:overlap_eta}
\end{equation}
showing that the renormalization factor is the finite-temperature
generalization of the overlap between the displaced bath states. Consequently,
$\eta\rightarrow1$ in the weak-coupling and low-temperature limits,
recovering the bare tunneling amplitude $\Delta$, whereas $\eta\rightarrow0$
for strong coupling or high temperatures, corresponding to the suppression
of coherent tunneling and the onset of dynamical localization \citep{Criteria_for_small_polarion}.

Because the bath is in a thermal Gaussian state, Wick's theorem allows
the bath correlation functions generated by $H_{sb}'$ to be expressed
in terms of first- and second-order moments \citep{Breuer02}. By
construction, the bath operators in Eqs. (\ref{eq:B_1})-(\ref{eq:B_3})
satisfy $\left\langle B_{i}\right\rangle _{H_{b}}=0$ and hence $\left\langle H_{sb}'\right\rangle _{H_{b}}=0$
For $B_{2}$ and $B_{3}$, this follows from the vanishing of thermal
averages containing an odd number of bosonic creation and annihilation
operators, while the subtraction of $\eta$ in $B_{1}$ ensures the
same condition for $B_{1}$. This zero-mean property is important
for the perturbative treatment, since it eliminates all odd-order
contributions from the expansion in $H_{sb}'$. Consequently, when
the master equation is truncated at second order, the leading neglected
contributions are of fourth order. Without the condition $\left\langle H_{sb}'\right\rangle _{H_{b}}=0$,
third-order terms would remain in the expansion.

The variational parameters $\left\{ \xi_{k}\right\} $ are determined
by minimizing the Feynman-Bogoliubov upper bound on the free energy
\citep{S-H_variational,VP_Master_Eq_DC_and_Imhomo}. To first order
in $H_{sb}'$, the bound is
\begin{equation}
A_{B}=-\frac{1}{\beta}\ln\left[\textrm{tr}_{T}\left(e^{-\beta H_{0}'}\right)\right]+\left\langle H_{sb}'\right\rangle _{H_{0}'},\label{Free energy}
\end{equation}
where $H_{0}'=H_{s}'+H_{b}$. Since $\left\langle H_{sb}'\right\rangle _{H_{b}}=0$,
the second term vanishes, and minimization of $A_{B}$ yields \citep{S-H_variational,VP_Master_Eq_DC_and_Imhomo}:
\begin{equation}
\xi_{k}=\left[1+\frac{X}{\omega_{k}}\coth\left(\frac{\beta\omega_{k}}{2}\right)\tanh\left(\beta W\right)\right]^{-1},\label{xi_k_DC}
\end{equation}
where $W=\sqrt{\left(\eta\Delta/2\right)^{2}+\left(\varepsilon/2\right)^{2}}$
and $X=\left(\eta\Delta\right)^{2}/\left(2W\right)$. Because $\eta$
itself depends on $\left\{ \xi_{k}\right\} $, Eqs. (\ref{eta(xi_k)})
and (\ref{xi_k_DC}) must be solved self-consistently.

The variational transformation interpolates between the weak-coupling
and full-polaron limits. In the limit of vanishing tunneling, $\Delta\rightarrow0$
, or vanishing system-bath coupling, $g_{k}\rightarrow0$, the interaction
Hamiltonian $H_{sb}'$ vanishes, and the resulting models are exactly
solvable. The variational parameters satisfy $0\leq\xi_{k}\leq1$,
approaching $\xi_{k}\rightarrow1$ in the strong-coupling or weak-tunneling
limit, corresponding to the full-polaron transformation, and $\xi_{k}\rightarrow0$
in the weak-coupling limit, corresponding to an untransformed Hamiltonian.
The variational transformation therefore provides a continuous interpolation
between these two regimes.

Using the spectral density $J(\omega)=\sum_{k}|g_{k}|^{2}\delta(\omega-\omega_{k})$,
Eq. (\ref{eta(xi_k)}) becomes, in the continuum limit,
\begin{equation}
\eta=\exp\left[-2\int_{0}^{\infty}\frac{J\left(\omega\right)}{\omega^{2}}F\left(\omega\right)^{2}\coth\left(\frac{\beta\omega}{2}\right)d\omega\right],\label{eta(w)}
\end{equation}
where $F(\omega)=\xi_{k}$ for $\omega=\omega_{k}$, with $F(\omega)$
determined by Eq.~(\ref{xi_k_DC}). Equation (\ref{eta(w)}) must
therefore be solved self-consistently together with Eq. (\ref{xi_k_DC}).

In this work, we consider a super-Ohmic spectral density,
\begin{equation}
J\left(\omega\right)=\alpha\omega^{3}e^{-\omega/\omega_{c}},\label{eq:J(w)}
\end{equation}
where $\alpha$ characterizes the system-bath coupling strength and
$\omega_{c}$ is the bath cutoff frequency. This completes the construction
of the variational polaron Hamiltonian used in the subsequent derivation
of the non-Markovian master equation.

\subsection{Transformed initial state}

\label{sec:transformed_initial_state}

We consider an initially factorized total state of the form $\rho_{T}\left(0\right)=\rho_{s}\left(0\right)\otimes\rho_{b}$
where $\rho_{s}\left(0\right)$ denotes an arbitrary initial state
of the two-level system and $\rho_{b}$ is the thermal equilibrium
state of the bath. To obtain the initial state in the transformed
frame, we express the system density matrix as
\begin{equation}
\rho_{s}\left(0\right)=\begin{pmatrix}\rho_{11} & \rho_{10}\\
\rho_{01} & \rho_{00}
\end{pmatrix},\label{eq:rhos(0)}
\end{equation}
where $\rho_{\mu\nu}$ are the matrix elements of $\rho_{s}\left(0\right)$.
Applying the variational polaron transformation to $\rho_{T}\left(0\right)$,
the transformed initial state can be decomposed into four system-bath
components,
\begin{equation}
\rho_{T}'\left(0\right)=e^{V}\left(\rho_{s}(0)\otimes\rho_{b}\right)e^{-V}=\sum_{u=1}^{4}\rho_{s,\left(u\right)}'\otimes\rho_{b,\left(u\right)}^{\textrm{d}},\label{eq:displaced initial state}
\end{equation}
where the corresponding system components are
\begin{eqnarray}
\rho_{s,\left(1\right)}' & = & \rho_{11}\left|1\right\rangle \left\langle 1\right|,\label{eq:rhos1_p}\\
\rho_{s,\left(2\right)}' & = & \eta\rho_{01}\left|0\right\rangle \left\langle 1\right|,\label{eq:rhos2_p}\\
\rho_{s,\left(3\right)}' & = & \eta\rho_{10}\left|1\right\rangle \left\langle 0\right|,\label{eq:rhos3_p}\\
\rho_{s,\left(4\right)}' & = & \rho_{00}\left|0\right\rangle \left\langle 0\right|.\label{eq:rhos4_p}
\end{eqnarray}
The associated displaced bath states are
\begin{eqnarray}
\rho_{b,\left(1\right)}^{\textrm{d}} & = & e^{\frac{\Phi}{2}}\rho_{b}e^{-\frac{\Phi}{2}},\label{eq:displaced rhob1}\\
\rho_{b,\left(2\right)}^{\textrm{d}} & = & \eta^{-1}e^{-\frac{\Phi}{2}}\rho_{b}e^{-\frac{\Phi}{2}},\label{eq:displaced rhob2}\\
\rho_{b,\left(3\right)}^{\textrm{d}} & = & \eta^{-1}e^{\frac{\Phi}{2}}\rho_{b}e^{\frac{\Phi}{2}},\label{eq:displaced rhob3}\\
\rho_{b,\left(4\right)}^{\textrm{d}} & = & e^{-\frac{\Phi}{2}}\rho_{b}e^{\frac{\Phi}{2}},\label{eq:displaced rhob4}
\end{eqnarray}
where the normalization factors in Eqs. (\ref{eq:displaced rhob2})
and (\ref{eq:displaced rhob3}) follow from the thermal average of
the bath displacement operators. In particular, the normalization
factor is determined by the property given in Eq. (\ref{eq:overlap_eta})
and is identical to the renormalization factor $\eta$ introduced
in Eq. (\ref{eta(xi_k)}). Consequently, each $\rho_{b,\left(u\right)}^{\textrm{d}}$
is normalized, and Eq. (\ref{eq:displaced initial state}) provides
a decomposition of the transformed initial state into normalized bath
components.

Using the projection operators $P$ and $Q$ introduced above, the
transformed initial state can then be decomposed as
\begin{equation}
\rho_{T}'\left(0\right)=\rho_{s}'\left(0\right)\otimes\rho_{b}+Q\rho_{T}'\left(0\right),\label{eq:rho'_T(0)}
\end{equation}
where $P\rho_{T}'\left(0\right)=\textrm{tr}_{b}[\rho_{T}'\left(0\right)]\otimes\rho_{b}=\rho_{s}'\left(0\right)\otimes\rho_{b}$
and $Q\rho_{T}'\left(0\right)=\left(\mathcal{I}-P\right)\rho_{T}'\left(0\right)$.
It follows from Eqs. (\ref{eq:rhos1_p})-(\ref{eq:rhos4_p}) that
the reduced system state in the transformed frame is
\[
\rho_{s}'\left(0\right)=\begin{pmatrix}\rho_{11} & \eta\rho_{10}\\
\eta\rho_{01} & \rho_{00}
\end{pmatrix}.
\]
Thus, the polaron transformation modifies the initial system coherences
through the renormalization factor $\eta$, while leaving the populations
unchanged. In general, therefore, $\rho{}_{s}'\left(0\right)\neq\rho{}_{s}\left(0\right)$
when the initial state contains nonzero coherences, i.e., $\rho_{01}\neq0$
and $\rho_{10}\neq0$. In contrast, for an initially diagonal state,
$\rho_{01}=\rho_{10}=0$, one obtains $\rho{}_{s}'\left(0\right)=\rho{}_{s}\left(0\right)$.
More importantly, even when the original state is factorized, the
transformed state generally contains system-bath correlations. These
correlations are represented by the nonvanishing $Q$-projected component
$Q\rho_{T}'\left(0\right)=\rho_{T}'\left(0\right)-\rho_{s}'\left(0\right)\otimes\rho_{b}$.
This term provides the initial-state contribution to the inhomogeneous
part of the variational polaron master equation. Therefore, unlike
the weak-coupling master equation in Eq. (\ref{weak coupling master eq}),
where the initial state is taken to be factorized with respect to
the system and the thermal bath in the working frame, the transformed
initial state generally generates a nonzero inhomogeneous term through
the $Q$-projected component.

\subsection{Variational polaron master equation}

\label{sec:Variational_polaron_master_equation_driven}

Once $\eta$ is determined numerically, we derive a quantum master
equation by treating the transformed system-bath interaction $H_{sb}'$
to second order. In the variational-polaron-interaction picture, the
Liouville equation is
\[
\frac{\partial}{\partial t}\tilde{\rho}_{T}'\left(t\right)=-i\left[\tilde{H}_{sb}'\left(t\right),\tilde{\rho}_{T}'\left(t\right)\right].
\]
Applying the projection operators $P$ and $Q=\mathcal{I}-P$, and
tracing over the bath degrees of freedom, gives
\begin{equation}
\frac{\partial}{\partial t}\tilde{\rho}_{s}'\left(t\right)=-i\textrm{tr}_{b}\left[\tilde{H}_{sb}'\left(t\right),\tilde{\rho}_{T}'\left(t\right)\right]=\textrm{tr}_{b}\left[\tilde{\mathscr{L}}_{sb}'\left(t\right)Q\tilde{\rho}_{T}'\left(t\right)\right],\label{d_dt P rho_T(t) inter pic}
\end{equation}
where $\tilde{\mathscr{L}}_{sb}'\left(t\right)A=-i\left[\tilde{H}_{sb}'\left(t\right),A\right]$,
$\tilde{H}_{sb}'\left(t\right)=e^{iH_{0}'t}H_{sb}'e^{-iH_{0}'t}$,
and $\tilde{\rho}_{s}'\left(t\right)=\textrm{tr}_{b}\left[\tilde{\rho}_{T}'\left(t\right)\right]$.
Here, $H_{0}'$ is the unperturbed Hamiltonian in the variational
polaron frame, as previously introduced in Sec. \ref{sec:Variational-polaron_transformation}
{[}see Eq. (\ref{Free energy}){]}.

Following the projection-operator procedure used for the weak-coupling
master equation, we expand $Q\tilde{\rho}_{T}'\left(t\right)$ to
first order in $\tilde{H}_{sb}'$. Because the transformed interaction
contains multiple bath operators $B_{i}$, and because $Q\rho_{T}'\left(0\right)$
can be nonzero for a general transformed initial state, the resulting
equation contains both homogeneous and inhomogeneous contributions.
After tracing over the bath degrees of freedom and transforming back
to the variational polaron frame, we obtain the second-order time-local
master equation
\begin{eqnarray}
\frac{\partial}{\partial t}\rho_{s}'\left(t\right) & = & \mathscr{L}_{s}'\rho_{s}'\left(t\right)\nonumber \\
 &  & +\sum_{i,j}\mathscr{L}_{i}\left[D_{ij}\left(t\right)\rho_{s}'\left(t\right)+\textrm{H.c.}\right]\nonumber \\
 &  & +\sum_{i,j,u}\mathscr{L}_{i}\left\{ C_{i}^{\textrm{d},\left(u\right)}\left(t\right)\mathcal{G}_{s}'\left(t,0\right)\rho_{s,\left(u\right)}'\right.\nonumber \\
 &  & \left.+\left[D_{ij}^{\textrm{d},\left(u\right)}\left(t\right)\mathcal{G}_{s}'\left(t,0\right)\rho_{s,\left(u\right)}'+\textrm{H.c.}\right]\right\} \label{VP master eq}
\end{eqnarray}
where $\rho_{s}'\left(t\right)=\textrm{tr}_{b}\left[e^{V}\rho_{T}\left(t\right)e^{-V}\right]$
describes the reduced system dynamics in the variational polaron frame.
The first two lines of Eq. (\ref{VP master eq}) describe the homogeneous
evolution generated by the transformed system Hamiltonian and the
second-order system-bath interaction. The last two lines contain the
inhomogeneous contributions originating from the $Q$-projected part
of the transformed initial state.

The second-order kernels are defined as
\begin{equation}
D_{ij}\left(t\right)=-i\int_{0}^{t}C_{ij}\left(t-t'\right)\mathcal{G}_{s}'\left(t,t'\right)\varLambda_{j}dt',\label{D_ij(t)}
\end{equation}
and
\begin{equation}
D_{ij}^{\textrm{d},\left(u\right)}\left(t\right)=-i\int_{0}^{t}C_{ij}^{\textrm{d},\left(u\right)}\left(t,t'\right)\mathcal{G}_{s}'\left(t,t'\right)\varLambda_{j}dt'.\label{eq:Dij^d(t,t')}
\end{equation}
Here, $i,j=1,2,3,$ and $\mathscr{L}_{i}A=-i\left[\varLambda_{i},A\right]$.
The propagator $\mathcal{G}_{s}'\left(t,t'\right)$ is generated by
the time-independent transformed system Hamiltonian and is given by
$\mathcal{G}_{s}'\left(t,t'\right)=\exp\left[\mathscr{L}_{s}'\left(t-t'\right)\right]$,
with $\mathscr{L}_{s}'A=-i\left[H_{s}',A\right]$, it can be evaluated
independently from $\dot{\mathcal{G}}_{s}'\left(t,t'\right)=\mathscr{L}_{s}'\mathcal{G}_{s}'\left(t,t'\right)$
with the initial condition $\mathcal{G}_{s}'\left(t',t'\right)=I$,
where $I$ denotes the identity operator. 

The homogeneous bath correlation functions entering Eq. (\ref{D_ij(t)})
are defined by
\[
C_{ij}\left(\tau\right)\equiv\left\langle B_{i}B_{j}\left(-\tau\right)\right\rangle _{H_{b}}=\textrm{tr}_{b}\left[B_{i}e^{-iH_{b}\tau}B_{j}e^{iH_{b}\tau}\rho_{b}\right].
\]
For the transformed interaction considered here, the nonvanishing
correlation functions are \citep{VP_Master_Eq_DC_and_Imhomo}
\begin{eqnarray}
C_{11}\left(\tau\right) & = & \eta^{2}\left\{ \cosh\left[r\left(\tau\right)\right]-1\right\} ,\label{eq:C11}\\
C_{22}\left(\tau\right) & = & -\eta^{2}\sinh\left[r\left(\tau\right)\right],\label{eq:C22}\\
C_{33}\left(\tau\right) & = & \int_{0}^{\infty}d\omega J\left(\omega\right)\left[1-F\left(\omega\right)\right]^{2}\nonumber \\
 &  & \times\left[\cos\left(\omega\tau\right)\coth\left(\frac{\beta\omega}{2}\right)-i\sin\left(\omega\tau\right)\right],\label{C33(tau)}\\
C_{32}\left(\tau\right) & = & 2\eta\int_{0}^{\infty}d\omega\frac{J\left(\omega\right)}{\omega}F\left(\omega\right)\left[1-F\left(\omega\right)\right]\nonumber \\
 &  & \times\left[\sin\left(\omega\tau\right)\coth\left(\frac{\beta\omega}{2}\right)+i\cos\left(\omega\tau\right)\right],\label{C32(tau)}
\end{eqnarray}
and $C_{23}\left(\tau\right)=-C_{32}\left(\tau\right)$, where
\begin{eqnarray}
r\left(\tau\right) & = & -4\int_{0}^{\infty}d\omega\frac{J\left(\omega\right)}{\omega^{2}}\left[F\left(\omega\right)\right]^{2}\nonumber \\
 &  & \times\left[\cos\left(\omega\tau\right)\coth\left(\frac{\beta\omega}{2}\right)-i\sin\left(\omega\tau\right)\right].\label{r(tau)}
\end{eqnarray}
All other bath correlation functions vanish, i.e., $C_{13}\left(\tau\right)=C_{31}\left(\tau\right)=C_{12}\left(\tau\right)=C_{21}\left(\tau\right)=0$. 

The functions $C_{i}^{\textrm{d},\left(u\right)}\left(t\right)$ and
$C_{ij}^{\textrm{d},\left(u\right)}\left(t,t'\right)$ in the inhomogeneous
terms are determined by the $Q$-projected part of the transformed
initial state. Their explicit definitions, together with the relations
used to reduce them to a smaller set of independent functions, are
given in Appendix \ref{sec:appendixA}. In particular, the Hermiticity
of the bath operators and the cyclic property of the trace allow the
required first- and second-order initial-state-induced functions to
be expressed in terms of the basic functions introduced there.

This completes the derivation of the master equation. The numerical
treatment of Eq. (\ref{VP master eq}), including the exponential
decomposition of the bath correlation functions and the conversion
of the integral kernels into a closed set of differential equations,
is described next.

To efficiently solve Eq.~(\ref{VP master eq}), we express each
bath correlation function as a sum of exponential terms,
\begin{equation}
C_{ij}\left(\tau\right)=\sum_{m}\alpha_{ij,m}e^{\gamma_{ij,m}\tau},\label{C_m,ij VP}
\end{equation}
wherein $\alpha_{ij,m}$ and $\gamma_{ij,m}$ are complex coefficients
obtained from numerical methods. Substituting Eq.~(\ref{C_m,ij VP})
into Eq.~(\ref{D_ij(t)}) gives $D_{ij}\left(t\right)=\sum_{m}D_{ij,m}\left(t\right)$,
where 
\begin{equation}
D_{ij,m}\left(t\right)=-i\int_{0}^{t}\alpha_{ij,m}e^{\gamma_{ij,m}\tau}\mathcal{G}_{s}'\left(t,t'\right)\varLambda_{j}dt',\label{D_ij,m(t)}
\end{equation}
and $\tau=t-t'$.

The initial-state-induced kernels $D_{ij}^{\textrm{d},\left(u\right)}\left(t\right)$
can likewise be expressed in terms of the auxiliary functions $D_{ij,m}\left(t\right)$
and the functions defined by
\begin{equation}
D_{l,ij,m,n}^{\textrm{d}_{q}}\left(t\right)=-i\int_{0}^{t}\eta^{-1}C_{l}^{\textrm{d}_{q}}\left(t'\right)\alpha_{ij,m}e^{\gamma_{ij,m}\tau}\mathcal{G}_{s}'\left(t,t'\right)\varLambda_{n}dt',\label{D^dq_l,ij,m,n(t)}
\end{equation}
and
\begin{equation}
D_{i}^{\textrm{d}_{q}}\left(t\right)=-i\int_{0}^{t}C_{i}^{\textrm{d}_{q}}\left(t'\right)\mathcal{G}_{s}'\left(t,t'\right)\varLambda_{i}dt',\label{D^dq_i(t)}
\end{equation}
where $l,i,j,n\in\left\{ 1,2,3\right\} $. The functions $C_{l}^{\textrm{d}_{q}}\left(t\right)$
arise from bath terms involving the bath operator $B_{l}$ associated
with the diagonal ($q=1$) and off-diagonal ($q=2$) components of
the $Q$-projected transformed initial state, and they are related
to $C_{i}^{\textrm{d},\left(u\right)}\left(t\right)$, with details
given in Appendix \ref{sec:appendixA}.

To solve Eq.~(\ref{VP master eq}), we therefore need to calculate
the three types of integral functions $D_{ij,m}\left(t\right)$, $D_{l,ij,m,n}^{\textrm{d}_{q}}\left(t\right)$
and $D_{i}^{\textrm{d}_{q}}\left(t\right)$. Their time derivatives
are given by{\small{}
\begin{align}
\dot{D}_{ij,m}\left(t\right) & =\left[\mathscr{L}_{s}'+\gamma_{ij,m}\right]D_{ij,m}\left(t\right)-i\alpha_{ij,m}\varLambda_{j},\label{dot_D_ij,m(t)}\\
\dot{D}_{l,ij,m,n}^{\textrm{d}_{q}}\left(t\right) & =\left[\mathscr{L}_{s}'+\gamma_{ij,m}\right]D_{l,ij,m,n}^{\textrm{d}_{q}}\left(t\right)-i\eta^{-1}C_{l}^{\textrm{d}_{q}}\left(t\right)\alpha_{ij,m}\varLambda_{n},\label{dot D^dq_l,ij,m,n(t)}\\
\dot{D}_{i}^{\textrm{d}_{q}}\left(t\right) & =\mathscr{L}_{s}'D_{i}^{\textrm{d}_{q}}\left(t\right)-iC_{i}^{\textrm{d}_{q}}\left(t\right)\varLambda_{i}.\label{dot_D_^dq_i(t)}
\end{align}
}Equations (\ref{dot_D_ij,m(t)})-(\ref{dot_D_^dq_i(t)}) together
with Eq.~(\ref{VP master eq}), form a set of differential equations
for our variational polaron master equation approach. Equations (\ref{dot D^dq_l,ij,m,n(t)})
and (\ref{dot_D_^dq_i(t)}) describe the bath relaxation effects induced
by the transformed initial state through the $Q$-projected part,
which contribute to the inhomogeneous terms of the variational polaron
master equation. This formulation converts the time-integral kernels
into differential equations and provides an efficient numerical implementation
of the non-Markovian master equation.

\section{Expectation values}

\label{sec:expectation_value}

In this section, we consider the evaluation of system observables
under the variational polaron transformation while retaining the inhomogeneous
terms. Since the system operator in $V$ is $\sigma_{z}$, the transformation
commutes with $\sigma_{z}$, such that $\sigma_{z}'=e^{V}\sigma_{z}e^{-V}=\sigma_{z}$.
Consequently, the population
\begin{equation}
P\left(t\right)=\frac{1+\left\langle \sigma_{z}\left(t\right)\right\rangle }{2}=\left\langle 1\right|\rho_{s}'\left(t\right)\left|1\right\rangle ,\label{eq:P(t)}
\end{equation}
or, equivalently, the population difference $\left\langle \sigma_{z}\left(t\right)\right\rangle =\textrm{tr}_{T}\left[\sigma_{z}'\rho_{T}'\left(t\right)\right]=\textrm{tr}_{s}\left[\sigma_{z}\rho_{s}'\left(t\right)\right]$,
can be obtained directly from the reduced system density matrix $\rho_{s}'\left(t\right)=\textrm{tr}_{b}\left[\rho_{T}'\left(t\right)\right]$,
where $\rho_{T}'\left(t\right)=e^{V}\rho_{T}\left(t\right)e^{-V}$.
Thus, the population dynamics can be directly evaluated using Eq.
\eqref{VP master eq}.

In contrast, the observables $\sigma_{x}$ and $\sigma_{y}$ do not
commute with the transformation and are transformed into system-bath
operators, $\sigma_{x}'=e^{V}\sigma_{x}e^{-V}$and $\sigma_{y}'=e^{V}\sigma_{y}e^{-V}$,
respectively. Their expectation values, $\left\langle \sigma_{x}\left(t\right)\right\rangle =\textrm{tr}_{T}\left[\sigma_{x}'\rho_{T}'\left(t\right)\right]$
and $\left\langle \sigma_{y}\left(t\right)\right\rangle =\textrm{tr}_{T}\left[\sigma_{y}'\rho_{T}'\left(t\right)\right]$
therefore cannot be obtained directly from $\rho_{s}'\left(t\right)$.
To address this issue, we consider the annihilation operator $\sigma_{-}=(\sigma_{x}-i\sigma_{y})/2$,
and decompose its expectation value according to the $P$- and $Q$-projected
components as
\begin{equation}
\left\langle \sigma_{-}\left(t\right)\right\rangle =\textrm{tr}_{T}\left[\sigma_{-}'\left(P+Q\right)\rho_{T}'\left(t\right)\right]=\left\langle \sigma_{-}\left(t\right)\right\rangle _{P}+\left\langle \sigma_{-}\left(t\right)\right\rangle _{Q}.\label{eq:<sigma_m(t)>}
\end{equation}
Under the transformation,
\[
\sigma_{-}'=e^{V}\sigma_{-}e^{-V}=\sigma_{-}\left(\eta+B_{1}+iB_{2}\right),
\]
where $B_{1}$ and $B_{2}$ are defined in Eqs. \eqref{eq:B_1} and
\eqref{eq:B_2}. The terms involving $B_{1}$ and $B_{2}$ do not
contribute to the $P$-projected part because $\textrm{tr}_{b}\left(B_{1}\rho_{b}\right)=\textrm{tr}_{b}\left(B_{2}\rho_{b}\right)=0$.
It therefore follows that
\begin{equation}
\left\langle \sigma_{-}\left(t\right)\right\rangle _{P}=\textrm{tr}_{s}\left\{ \textrm{tr}_{b}\left[\sigma_{-}'P\rho_{T}'\left(t\right)\right]\right\} =\textrm{tr}_{s}\left[\eta\sigma_{-}\rho{}_{s}'\left(t\right)\right],\label{eq:sigma_P_interact_pic}
\end{equation}
which can be evaluated directly from the reduced system density matrix.

The $Q$-projected contribution,
\begin{eqnarray}
\left\langle \sigma_{-}\left(t\right)\right\rangle _{Q} & = & \textrm{tr}_{s}\left\{ \textrm{tr}_{b}\left[\sigma_{-}'Q\rho_{T}'\left(t\right)\right]\right\} ,\label{eq:sigma_Q}
\end{eqnarray}
cannot be obtained directly from $\rho{}_{s}'\left(t\right)$, because
$\sigma_{-}'$ contains bath operators. In principle, this contribution
can be evaluated by expanding $Q\rho_{T}'\left(t\right)$ to first
order in $H_{sb}'$, as in previous treatments \citep{multisite_SP_time-local_initial_fac_in_imhomo,off_diagonal_expectaion_sp_2012}.
Here, however, we exploit the similarity between this calculation
and the derivation of the inhomogeneous terms in Eq. \eqref{VP master eq},
which allows the result to be obtained using the same projection-operator
framework.

First, the $P$-projected contribution in Eq. \eqref{eq:sigma_P_interact_pic}
has the same structure as the first term of Eq. \eqref{VP master eq},
with the system Liouvillian $\mathscr{L}_{s}'$ replaced by the system
operator $\eta\sigma_{-}$. Second, in the interaction picture with
respect to $H_{0}'=H_{s}'+H_{b}$, where $\tilde{\sigma}_{-}'\left(t\right)=e^{iH_{0}'t}\sigma_{-}'e^{-iH_{0}'t}$
and $\tilde{\rho}_{T}'\left(t\right)=e^{iH_{0}'t}\rho_{T}'\left(t\right)e^{-iH_{0}'t}$,
the terms inside the bath trace in Eq. \eqref{d_dt P rho_T(t) inter pic}
and Eq. \eqref{eq:sigma_Q} have the same structure, with the interaction
Liouvillian $\tilde{\mathscr{L}}_{sb}'\left(t\right)$ replaced by
$\tilde{\sigma}_{-}'\left(t\right)$. Consequently, the expression
for $\left\langle \sigma_{-}\left(t\right)\right\rangle $ can be
obtained from Eq. \eqref{VP master eq} by making the replacements
$\mathscr{L}_{s}'\rightarrow\eta\sigma_{-}$, $\mathscr{L}_{1}\rightarrow\sigma_{-}$,
$\mathscr{L}_{2}\rightarrow i\sigma_{-}$ and $\mathscr{L}_{3}\rightarrow0$.
We thus obtain
\begin{equation}
\left\langle \sigma_{-}\left(t\right)\right\rangle =\textrm{tr}_{s}\left[\sigma_{-}\rho_{s,-}\left(t\right)\right],\label{eq:sigma_P_Q}
\end{equation}
where the modified density matrix $\rho_{s,-}\left(t\right)$ is given
by
\begin{align}
\rho_{s,-}\left(t\right) & =\eta\rho_{s}'\left(t\right)\nonumber \\
 & +\left[D_{11}\left(t\right)\rho_{s}'\left(t\right)+\textrm{H.c.}\right]\nonumber \\
 & +i\left\{ \left[D_{22}\left(t\right)+D_{23}\left(t\right)\right]\rho_{s}'\left(t\right)+\textrm{H.c.}\right\} \nonumber \\
 & +\sum_{j,u}\left(\left[C_{1}^{\textrm{d},\left(u\right)}\left(t\right)+iC_{2}^{\textrm{d},\left(u\right)}\left(t\right)\right]\mathcal{G}_{s}'\left(t,0\right)\rho_{s,\left(u\right)}'\right.\nonumber \\
 & \left.+\left\{ \left[D_{1j}^{\textrm{d},\left(u\right)}\left(t\right)+iD_{2j}^{\textrm{d},\left(u\right)}\left(t\right)\right]\mathcal{G}_{s}'\left(t,0\right)\rho_{s,\left(u\right)}'+\textrm{H.c.}\right\} \right).\label{eq:rho_s_}
\end{align}
Equation \eqref{eq:sigma_P_Q} shows that, although the reduced density
matrix $\rho_{s}'\left(t\right)$ contains the information required
to determine the population, the expectation value of $\sigma_{-}$
requires the modified density matrix $\rho_{s,-}\left(t\right)$ defined
in Eq. \eqref{eq:rho_s_}. In particular, the latter incorporates
both the $P$-projected and $Q$-projected contributions associated
with the transformed initial state.

In the pure-dephasing limit, we show in Appendix \ref{sec:appendix-pure_dephasing}
that Eq.~(\ref{eq:sigma_P_Q}) reduces to the exact analytical solution
\citep{pure_dephasing_analytical_solution1990,pure_dephasing_analytical_solution2012,Breuer02,pure_dephasing_analytical_solution_Palma1996},
\begin{eqnarray}
\left\langle \sigma_{-}\left(t\right)\right\rangle  & = & e^{-4\int_{0}^{\infty}d\omega\frac{J\left(\omega\right)}{\omega^{2}}\coth\left(\frac{\beta\omega}{2}\right)\left[1-\cos\left(\omega\tau\right)\right]}\rho_{10}e^{-i\epsilon t}.\label{eq:PD_analytical_solution}
\end{eqnarray}
This provides an alternative derivation of the analytical solution
of the pure-dephasing model using the (variational) polaron transformation.

Finally, for a general polaron-transformed initial state $\rho_{T}'\left(0\right)$,
both the $P$-projected and $Q$-projected contributions must be retained
to obtain the correct dynamics. Combining Eqs. \eqref{eq:P(t)} and
(\ref{eq:sigma_P_Q}), we can reconstruct the reduced system density
matrix in the original frame as
\begin{equation}
\rho_{s}\left(t\right)=\begin{pmatrix}P\left(t\right) & \left\langle \sigma_{-}\left(t\right)\right\rangle \\
\left\langle \sigma_{-}\left(t\right)\right\rangle ^{*} & 1-P\left(t\right)
\end{pmatrix}.\label{eq:rho_s(t)}
\end{equation}
Thus, the complete system dynamics can be obtained from the population
and the coherence calculated from Eqs. \eqref{eq:P(t)} and (\ref{eq:sigma_P_Q}),
respectively. This provides a description of the non-Markovian dynamics
of the spin-boson model over a broader parameter regime beyond weak
system-bath coupling.

\section{Initial state and the trace distance}

\label{sec:Initial_state_trace_distance}

Following the discussion in the previous section, we categorize our
numerical results into two cases: the expectation values of operators
that commute with $\sigma_{z}$ and those of operators that do not
commute with $\sigma_{z}$. We find that the effect of the initial
state depends on the observable under consideration. For a more effective
comparison, we therefore consider different initial states for the
two cases in the following numerical analyses. Specifically, we consider
three initial states in the original laboratory frame, denoted by
$\left|-\right\rangle _{y}$ , $\left|+\right\rangle _{x}$ and $\left|+\right\rangle _{z}$.
These states are eigenstates of the corresponding Pauli operators
$\sigma_{y}$, $\sigma_{x}$, and $\sigma_{z}$, respectively. In
particular, $\left|+\right\rangle _{z}$ is equivalent to the state
$\left|1\right\rangle $, with eigenvalue $+1$.

As discussed in Sec. \ref{sec:transformed_initial_state}, obtaining
the corresponding variational-polaron-transformed initial states can
be computationally demanding. It is therefore natural to consider
simpler approximations to the transformed initial state and quantify
the resulting errors. Two simplified forms are particularly relevant.
The motivation for considering these forms is that the transformed
Hamiltonian in Eq. (\ref{H'_T}) retains the bath Hamiltonian $H_{b}$
in the same form as that in the original Hamiltonian in Eq. (\ref{H_T}).
One might therefore be tempted to retain the bath state $\rho_{b}$
in its original thermal form after the transformation.

For an initial system state $\left|\pm\right\rangle _{z}$, or more
generally a mixed state that commutes with $\sigma_{z}$, the reduced
system state satisfies $\rho_{s}'\left(0\right)=\rho_{s}\left(0\right)$,
as discussed in Sec. \ref{sec:transformed_initial_state}. In this
case, one may consider $\rho_{s}\left(0\right)\otimes\rho_{b}$ as
a simplified initial state. In contrast, for an initial system state
that is a superposition state, or more generally a mixed state that
does not commute with $\sigma_{z}$, one has $\rho_{s}'\left(0\right)\neq\rho{}_{s}\left(0\right)$.
In this case, another natural simplified form is $\rho_{s}'\left(0\right)\otimes\rho_{b}$,
in which the transformed reduced system state is retained while the
bath state is assumed to remain thermal. These two choices help explain
the trends observed in the subsequent variational-polaron results:
a relatively smaller error is obtained using $\rho_{s}(0)\otimes\rho_{b}$
for the initial state $\left|+\right\rangle _{z}$, whereas $\rho_{s}'\left(0\right)\otimes\rho_{b}$
generally provides a better approximation for the initial states $\left|-\right\rangle _{y}$
or $\left|+\right\rangle _{x}$.

As summarized in Table \ref{table-Initial_A_B_C}, the transformed
total initial state $\rho_{T}'\left(0\right)$ given in Eq. (\ref{eq:displaced initial state})
is denoted as Initial-A. This state represents the exact transformed
total initial state. The first simplified form, $\rho_{s}'\left(0\right)\otimes\rho_{b}$,
is denoted as Initial-B. It corresponds to the $P$-projected part
of Initial-A, with the normalization factor $\eta$ included in $\rho_{s}'\left(0\right)$.
The second simplified form, $\rho_{s}\left(0\right)\otimes\rho_{b}$,
is denoted as Initial-C. This form corresponds to the original factorized
initial state, with the system and bath assumed to remain untransformed
in the polaron frame. Initial-A and Initial-B therefore have the same
reduced system state by construction. When either Initial-B or Initial-C
is adopted as the initial state, the variational polaron master equation
contains no inhomogeneous term. Furthermore, if the initial system
state in the original laboratory frame is $\left|\pm\right\rangle _{z}$,
or more generally a mixed state that commutes with $\sigma_{z}$,
Initial-B and Initial-C become identical. The following numerical
results demonstrate that the validity of these two simplified forms
depends on the relevant bath-relaxation timescale, which is reflected
in the qubit tunneling amplitude $\Delta$.

In studies based on the polaron transformation, comparisons have often
focused primarily on the dynamics of the populations. A more complete
characterization of the reduced system dynamics, however, requires
consideration of both diagonal and off-diagonal elements of the reduced
system density operator in the original frame. To this end, we introduce
the trace distance, which provides a sensitive measure for distinguishing
two quantum states $\rho^{\alpha}$ and $\rho^{\beta}$ \citep{multisite_SP_time-local_initial_fac_in_imhomo,initial_paper}:
\begin{equation}
{D}\left(\rho^{\alpha},\rho^{\beta}\right)=\frac{\left\Vert \rho^{\alpha}-\rho^{\beta}\right\Vert }{2}=\frac{1}{2}\textrm{tr}\sqrt{\left(\rho^{\alpha}-\rho^{\beta}\right)^{2}}.\label{eq:TD}
\end{equation}
The trace distance incorporates contributions from both the diagonal
and off-diagonal matrix elements of the reduced system density operator,
which are directly related to the dynamics of $P\left(t\right)$ and
$\left\langle \sigma_{-}\left(t\right)\right\rangle $, respectively.
For example, we take the time-evolved reduced system density matrix
$\rho_{s}\left(t\right)$ obtained from Eq. \eqref{eq:rho_s(t)} with
Initial-A as $\rho^{\alpha}$, and that obtained with Initial-B as
$\rho^{\beta}$. We denote the corresponding trace distance between
them by \textquotedblleft A \& B.\textquotedblright{} Similarly, \textquotedblleft A
\& C\textquotedblright{} denotes the trace distance between the reduced
system states obtained from Initial-A and Initial-C.

The trace distance \textquotedblleft A \& B\textquotedblright{} therefore
quantifies the effect of the $Q$-projected part of the variational-polaron-transformed
initial state, whereas \textquotedblleft A \& C\textquotedblright{}
quantifies the combined effects of the $P$- and $Q$-projected parts.
Since Initial-A and Initial-B have the same reduced system state by
construction, their initial trace distance is always zero. In contrast,
Initial-C does not contain the factor $\eta$ in the off-diagonal
components of the reduced system state, as shown in Eqs. (\ref{eq:rhos2_p})-(\ref{eq:rhos3_p}).
Consequently, the initial trace distance for \textquotedblleft A \&
C\textquotedblright{} is zero when $\rho{}_{s}\left(0\right)$ has
no off-diagonal elements and is generally nonzero when $\rho{}_{s}\left(0\right)$
has nonzero off-diagonal elements.

The nonequilibrium displaced bath states generated by the polaron
transformation, as given in Eqs. (\ref{eq:displaced rhob1})-(\ref{eq:displaced rhob4}),
are expected to relax toward the thermal equilibrium bath state on
the bath-relaxation timescale. Accordingly, both trace distances \textquotedblleft A
\& B\textquotedblright{} and \textquotedblleft A \& C\textquotedblright{}
will converge to zero in the long-time limit. The dynamics of these
trace distances therefore provide a direct measure of the effects
associated with bath relaxation induced by the transformed initial
states \citep{VP_Master_Eq_DC_and_Imhomo,coherent_initial_state_inhomo_SP}.

\begin{table}
\begin{tabular}{|c|c|c|}
\hline 
Treatment & Notation & Definition\tabularnewline
\hline 
Initial-A & $\rho_{T}'\left(0\right)$ & Full polaron-transformed state\tabularnewline
\hline 
Initial-B & $\rho_{s}'\left(0\right)\otimes\rho_{b}$ & $P$-projected component of Initial-A\tabularnewline
\hline 
Initial-C & $\rho_{s}\left(0\right)\otimes\rho_{b}$ & Untransformed factorized state\tabularnewline
\hline 
\end{tabular}

\caption{Notation and definitions of the polaron-transformed initial state
Initial-A and the two simplified initial-state treatments, Initial-B
and Initial-C, used in the variational polaron master equation approach.}

\label{table-Initial_A_B_C}
\end{table}

\section{Dynamics of the undriven spin-boson model}

\label{sec:Dynamics_result_undriven}

\subsection{Dynamics of the population $P\left(t\right)$}

To provide a clear description of the effects of the initial states,
we present numerical results under different parameter regimes by
varying the qubit tunneling amplitude $\Delta$ and temperature $T$,
while keeping the system-bath coupling strength characterized by the
coupling constant $\alpha$. These parameters allow us to examine
how the relative importance of the initial-state effects changes with
the system dynamics and the surrounding environment.

\subsubsection{Dependence on the qubit tunneling amplitude $\Delta$}

\begin{figure}
\includegraphics[width=1\columnwidth]{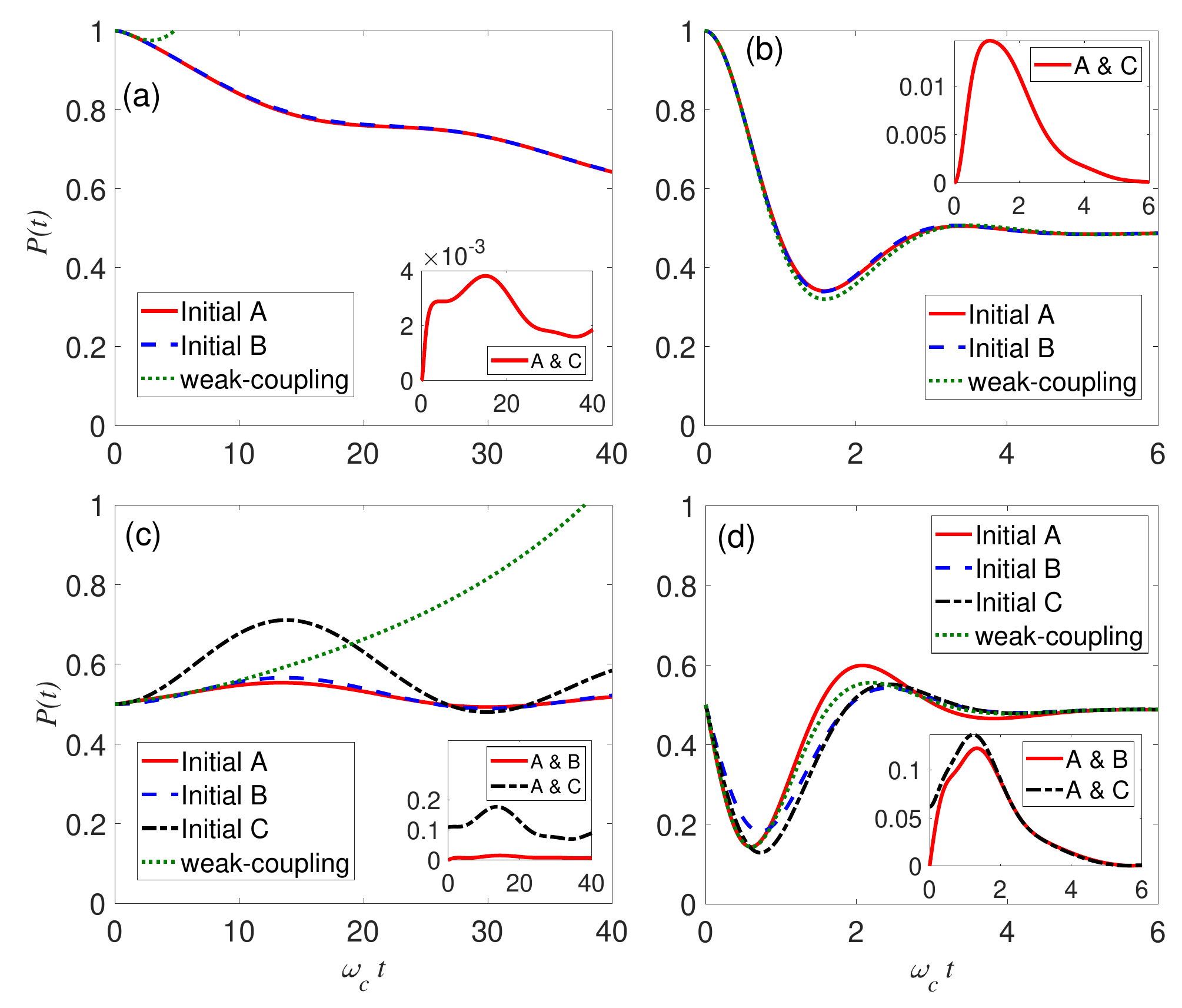}\caption{Dynamics of population $P\left(t\right)$ for the spin-boson model
calculated using the weak-coupling method (dark-green dotted line),
and variational polaron method with the inhomogeneous terms (Initial-A,
red solid line) and without the inhomogeneous terms (Initial-B, blue
dashed line; Initial-C, black dash-dotted line) . The initial states
and parameters are: (a) $|+\rangle_{z}$, $\Delta=0.2\omega_{c}$.
(b) $|+\rangle_{z}$, $\Delta=2\omega_{c}$. (c) $|+\rangle_{x}$,
$\Delta=0.2\omega_{c}$. (d) $|-\rangle_{y}$, $\Delta=2\omega_{c}$.
The corresponding insets show the time evolutions of the trace distances
between the reduced system states evolving from Initial-A and Initial-B,
denoted by ``A \& B'' (red solid line), and from Initial-A and Initial-C,
denoted by ``A \& C'' (black dash-dotted line). Other parameters
are $\alpha=0.5/\left(2\pi\omega_{c}^{2}\right)$, $\varepsilon=0.2\omega_{c}$,
and $k_{B}T=4\omega_{c}$.}
\label{fig:P(t)_spin-boson_T_4} 
\end{figure}

\begin{figure}
\includegraphics[width=1\columnwidth]{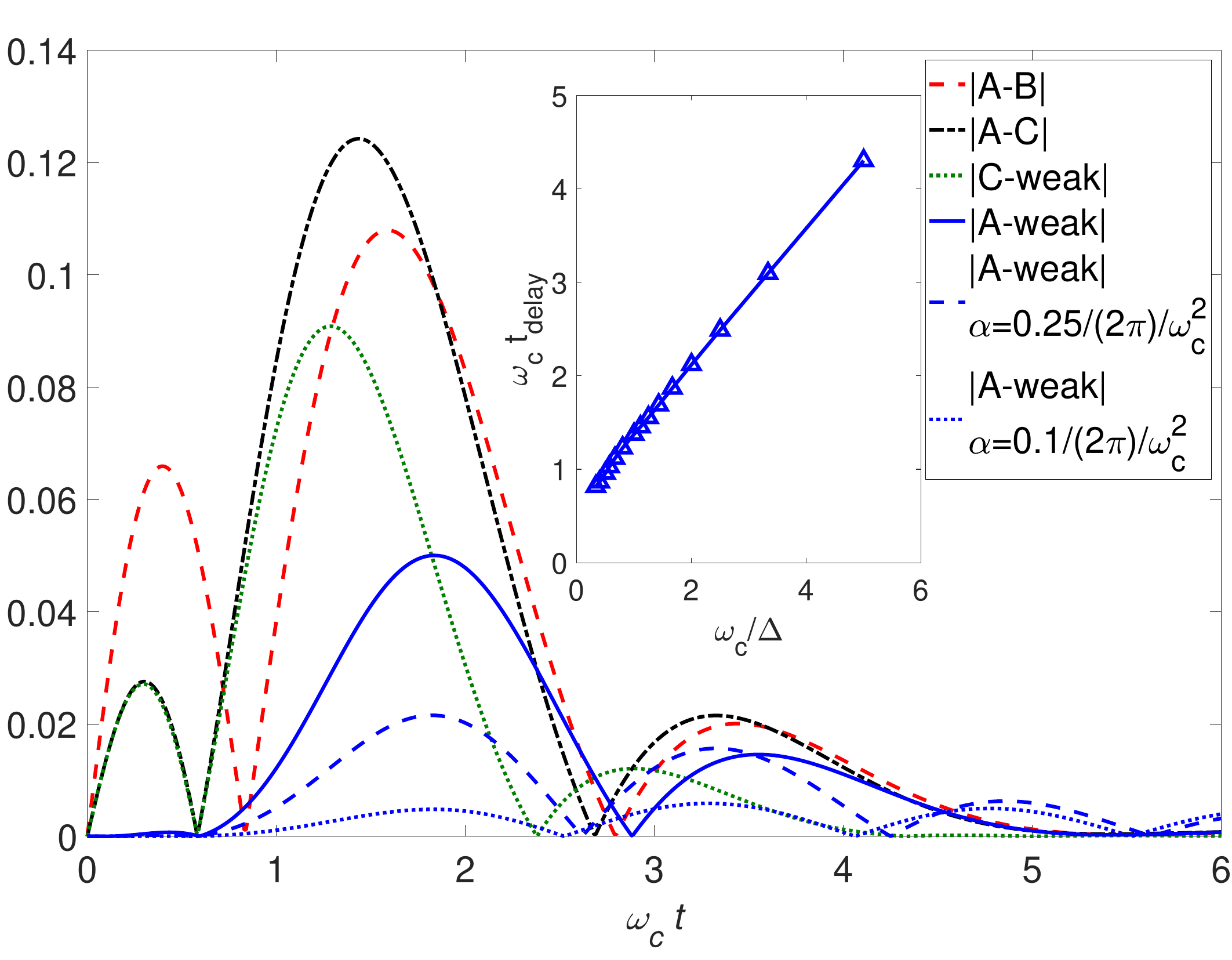} \caption{Absolute differences between the population dynamics $P\left(t\right)$
shown in Fig. \ref{fig:P(t)_spin-boson_T_4}(d). The blue solid line,
\textquotedblleft $\left|\textrm{A-weak}\right|$,\textquotedblright{}
denotes the difference between the variational polaron method with
Initial-A and the weak-coupling method. Similarly, \textquotedblleft $\left|\textrm{A-B}\right|$\textquotedblright{}
(red dashed line), \textquotedblleft $\left|\textrm{A-C}\right|$\textquotedblright{}
(black dash-dotted line), and \textquotedblleft $\left|\textrm{C-weak}\right|$\textquotedblright{}
(dark-green dotted line) denote the corresponding differences between
the indicated methods or initial-state treatments. The other parameters
are the same as those in Fig. \ref{fig:P(t)_spin-boson_T_4}(d). The
$\left|\textrm{A-weak}\right|$ curve exhibits a delay of approximately
$\omega_{c}t=1$ compared with $\left|\textrm{A-B}\right|$, $\left|\textrm{A-C}\right|$,
and $\left|\textrm{C-weak}\right|$. This delay may be associated
with multi-bath-quanta processes. The inset shows the dependence of
the delay time, defined as the time at which $\left|\textrm{A-weak}\right|$
reaches 0.01, on $\omega_{c}/\Delta$. In addition to $\alpha=0.5/\left(2\pi\omega_{c}^{2}\right)$
(blue solid line), as used in Fig. \ref{fig:P(t)_spin-boson_T_4}(d),
results for two smaller system-bath coupling strengths, $\alpha=0.25/\left(2\pi\omega_{c}^{2}\right)$
(blue dashed line) and $\alpha=0.1/\left(2\pi\omega_{c}^{2}\right)$
(blue dotted line), are shown, while the other parameters remain unchanged.}
\label{fig:multi-bath-quanta_delay} 
\end{figure}

We first investigate the time evolution of the population $P\left(t\right)$
and use the trace distance to provide a clearer comparison of the
dynamics, as shown in Fig. \ref{fig:P(t)_spin-boson_T_4}. We consider
two values of the qubit tunneling amplitude, $\Delta=0.2\omega_{c}$
{[}Figs. \ref{fig:P(t)_spin-boson_T_4}(a) and \ref{fig:P(t)_spin-boson_T_4}(c){]}
and $\Delta=2\omega_{c}$ {[}Figs. \ref{fig:P(t)_spin-boson_T_4}(b)
and \ref{fig:P(t)_spin-boson_T_4}(d){]}. The initial quantum state
is $|+\rangle_{z}$ in Figs. \ref{fig:P(t)_spin-boson_T_4}(a) and
\ref{fig:P(t)_spin-boson_T_4}(b), while $|+\rangle_{x}$ and $|-\rangle_{y}$
are considered in Figs. \ref{fig:P(t)_spin-boson_T_4}(c) and \ref{fig:P(t)_spin-boson_T_4}(d),
respectively. The results are obtained using the weak-coupling method
(green dotted line) and the variational polaron method with Initial-A
(red solid line), Initial-B (blue dashed line), and Initial-C (black
dash-dotted line).

For $\Delta=0.2\omega_{c}$, as shown in Figs. \ref{fig:P(t)_spin-boson_T_4}(a)
and \ref{fig:P(t)_spin-boson_T_4}(c), the weak-coupling result develops
unphysical population dynamics. In particular, the population of the
excited state $\left|1\right\rangle $ can exceed unity, while that
of the ground state $\left|0\right\rangle $ can become negative.
This behavior results from the relatively strong system-bath coupling,
$\alpha=0.50/\left(2\pi\omega_{c}^{2}\right)$, and the high temperature,
$k_{B}T=4\omega_{c}=20\Delta$, which place the parameters outside
the regime where the second-order weak-coupling master equation provides
a reliable description. The resulting violation of positivity is a
known limitation of second-order master equations \citep{Full_polaron_steady_state_2020}.
In contrast, for the larger tunneling amplitude $\Delta=2\omega_{c}$,
shown in Figs. \ref{fig:P(t)_spin-boson_T_4}(b) and \ref{fig:P(t)_spin-boson_T_4}(d),
the system is closer to the weak-coupling regime, and the weak-coupling
results become closer to those obtained from Initial-A. As shown in
the following subsection, the agreement is further improved at zero
temperature, where the weak-coupling and variational-polaron results
converge.

Because the differences among the population dynamics in Fig. \ref{fig:P(t)_spin-boson_T_4}
are not always readily discernible, we additionally show the trace
distances in the insets, particularly in Figs. \ref{fig:P(t)_spin-boson_T_4}(a)
and \ref{fig:P(t)_spin-boson_T_4}(b). When the original initial state
is $|+\rangle_{z}$, the dynamics obtained from Initial-A and Initial-B
are nearly identical. In contrast, when the original initial states
are $|+\rangle_{x}$ and $|-\rangle_{y}$, as shown in Figs. \ref{fig:P(t)_spin-boson_T_4}(c)
and \ref{fig:P(t)_spin-boson_T_4}(d), the effect of the variational
polaron transformation on the initial state becomes more pronounced,
as reflected in both $P\left(t\right)$ and the trace distance.

For the smaller tunneling amplitude $\Delta=0.2\omega_{c}$ in Fig.
\ref{fig:P(t)_spin-boson_T_4}(c), the population dynamics obtained
from Initial-A and Initial-C show more pronounced differences, while
the difference between Initial-A and Initial-B remains relatively
small. Correspondingly, the trace distance between Initial-A and Initial-C
is appreciable, whereas that between Initial-A and Initial-B remains
close to zero throughout the evolution. This indicates that, in this
parameter regime, the $P$-projected part of the transformed initial
state makes the dominant contribution to the dynamics. Initial-B therefore
provides a good approximation to the polaron-transformed initial state
in this regime. In contrast, the dynamics obtained from Initial-C
requires a longer time to approach that of Initial-A, reflecting the
additional effect associated with the transformation of the initial
state in the polaron frame.

For the larger tunneling amplitude $\Delta=2\omega_{c}$, the inset
of Fig. \ref{fig:P(t)_spin-boson_T_4}(d) shows a clear distinction
between the initial trace distances: the trace distance between Initial-A
and Initial-B is zero, whereas that between Initial-A and Initial-C
is nonzero. Thus, Initial-B and Initial-C represent distinct reduced
initial states in this case. During the subsequent evolution, the
two trace distances approach one another around $\omega_{c}t\sim2$
and eventually overlap, indicating that the reduced states associated
with Initial-B and Initial-C become indistinguishable. At a longer
time scale, $t\sim6\omega_{c}^{-1}$, which is much longer than the
characteristic qubit time scale $\sim1/\Delta$, both trace distances
tend to zero. Consequently, all three initial-state descriptions converge
to the same reduced state as the system approaches equilibrium with
its environment. The same behavior can also be observed in the population
dynamics shown in the main panel of Fig. \ref{fig:P(t)_spin-boson_T_4}(d).

The difference between Initial-B and Initial-C originates from the
factor $\eta$ appearing in Eqs. (\ref{eq:rhos2_p}) and (\ref{eq:rhos3_p}).
For a larger $\Delta$, the stronger coherent dynamics in the $\sigma_{x}$
direction induces more rapid Rabi oscillations, which drives the reduced
states associated with Initial-B and Initial-C toward one another.
Accordingly, compared with Fig. \ref{fig:P(t)_spin-boson_T_4}(c),
the Initial-C dynamics in Fig. \ref{fig:P(t)_spin-boson_T_4}(d) approaches
that of Initial-A on a shorter time scale before reaching equilibrium.

These results indicate that the $P$-projected part of Initial-A,
which describes the dressed initial state, has a significant influence
on the early-time dynamics. As the tunneling amplitude $\Delta$ increases,
however, the time interval over which this effect is prominent becomes
shorter. At later times, the influence of the $Q$-projected part
of Initial-A, which gives rise to the inhomogeneous terms in the master
equation, becomes increasingly apparent as the trace distances between
Initial-A and Initial-B and between Initial-A and Initial-C approach
and eventually overlap. In other words, the initial-state dependence
associated with the $P$-projected part becomes less significant at
longer times, allowing the contribution from the $Q$-projected part
to become more visible. The trace distance therefore provides a useful
measure for resolving these distinct initial-state effects, particularly
in the short- and intermediate-time dynamics.

\subsubsection{Delay time from the higher-order multi-bath-quanta process}

The results in Figs. \ref{fig:P(t)_spin-boson_T_4}(c) and \ref{fig:P(t)_spin-boson_T_4}(d)
show that the population $P\left(t\right)$ for Initial-B and Initial-C
begins to deviate from that for Initial-A immediately after $t=0$.
In contrast, the population obtained with Initial-A overlaps with
that obtained from the weak-coupling method during the initial stage
of the evolution. To characterize these differences and the initial
overlap, we extract the corresponding $P\left(t\right)$ data from
Fig. \ref{fig:P(t)_spin-boson_T_4}(d) and plot their absolute differences
in Fig. \ref{fig:multi-bath-quanta_delay}. We denote the difference
between Initial-A and the weak-coupling result by $\left|\textrm{A-weak}\right|$
(blue solid line), and similarly denote the differences between Initial-A
and Initial-B, Initial-A and Initial-C, and Initial-C and the weak-coupling
result by $\left|\textrm{A-B}\right|$ (red dashed line), $\left|\textrm{A-C}\right|$
(black dash-dotted line), and $\left|\textrm{C-weak}\right|$ (dark-green
dotted line), respectively. As shown in Fig. \ref{fig:multi-bath-quanta_delay},
$\left|\textrm{A-B}\right|$, $\left|\textrm{A-C}\right|$ and $\left|\textrm{C-weak}\right|$
increase immediately from $t=0$. In contrast, $\left|\textrm{A-weak}\right|$
exhibits a delay and remains close to zero during the initial stage
of the evolution, indicating an overlap between the population dynamics
obtained from Initial-A and the weak-coupling method.

As the system-bath coupling constant $\alpha$ decreases, the effect
of the initial state becomes weaker, and the population dynamics for
Initial-A, Initial-B, and Initial-C approach those obtained from the
weak-coupling method. This behavior is further illustrated in Fig.
\ref{fig:multi-bath-quanta_delay} by plotting $\left|\textrm{A-weak}\right|$
for two smaller values of $\alpha$, namely, $\alpha=0.25/\left(2\pi\omega_{c}^{2}\right)$
(blue dashed line) and $\alpha=0.1/\left(2\pi\omega_{c}^{2}\right)$
(blue dotted line). Although the magnitude of $\left|\textrm{A-weak}\right|$
becomes smaller as $\alpha$ decreases, a similar delay is still observed.

The population dynamics are primarily governed by the tunneling term
proportional to $\sigma_{x}$ in Eqs. (\ref{H_s}) and (\ref{H'_s})
for the weak-coupling method and Initial-A, respectively. The difference
between the two descriptions is characterized by the renormalization
factor $\eta$ {[}see Eq. (\ref{eta(xi_k)}){]}, which appears in
the renormalized tunneling term $\eta\Delta\sigma_{x}/2$ and incorporates
contributions from higher orders in the system-bath interaction \citep{Full_polaron_steady_state_2020}.
Within the polaron picture, the spin and the displaced bath oscillators
form a polaron state. A spin transition between the two levels is
therefore accompanied by a displacement of the bath oscillators, with
the renormalization factor $\eta$ accounting for the resulting multi-bath-quanta
processes and reducing the effective tunneling amplitude from $\Delta$
to $\eta\Delta$. The observed delay can thus be understood as a manifestation
of the higher-order multi-bath-quanta processes incorporated in the
variational polaron description. Since the characteristic timescale
associated with the tunneling dynamics is expected to scale approximately
as $1/\Delta$, a smaller $\Delta$ leads to a longer time interval
before the effect of these processes becomes appreciable. Their contribution
is also enhanced with increasing $\alpha$.

To quantify this timescale, we define the delay time $t_{\textrm{delay}}$
as the time at which $\left|\textrm{A-weak}\right|$ first reaches
0.01. This definition provides a measure of the time interval after
$t=0$ over which the population dynamics from Initial-A remain effectively
indistinguishable from those of the weak-coupling method according
to the chosen threshold. In the inset of Fig. \ref{fig:multi-bath-quanta_delay},
we plot $\omega_{c}t_{\textrm{delay}}$ as a function of $\omega_{c}/\Delta$,
with the other parameters fixed to those used in Fig. \ref{fig:P(t)_spin-boson_T_4}(d).
The approximately linear relation demonstrates that a smaller $\Delta$
corresponds to a longer delay time, consistent with the characteristic
timescale $t_{\textrm{delay}}\propto1/\Delta$. 

In the regime $\omega_{c}\geq\Delta\geq0.2\omega_{c}$, the weak-coupling
method breaks down and can even yield negative populations \citep{Full_polaron_steady_state_2020}.
Nevertheless, the delay time remains approximately linearly dependent
on $1/\Delta$, as shown in the inset of Fig. \ref{fig:multi-bath-quanta_delay}.
Thus, even when the weak-coupling method eventually becomes invalid,
it can still provide a good description of the population dynamics
during the initial time interval $0\leq t\apprle t_{\textrm{delay}}$.
For smaller values of $\Delta$, namely, $\Delta\leq0.1\omega_{c}$,
this linear relation is no longer observed under the given parameter
set. This regime approaches the pure-dephasing limit, where the weak-coupling
result becomes increasingly close to the result obtained with Initial-A.

In the pure-dephasing limit $\Delta\rightarrow0$, the system dynamics
can be solved exactly both by the variational polaron method with
Initial-A, as discussed above, and by the second-order weak-coupling
method \citep{pure_dephasing_2nd_mster_eq_2008}. In this limit, contributions
beyond second order in $H_{sb}$ vanish, and hence the higher-order
multi-bath-quanta processes responsible for the observed delay do
not contribute. Consequently, the population dynamics obtained from
Initial-A and the weak-coupling method are expected to overlap completely,
giving $\left|\textrm{A-weak}\right|=0$ at all times. 

We also note that no delay is observed in $\left|\textrm{C-weak}\right|$,
where Initial-C and the weak-coupling method correspond to different
physical initial states. In contrast, Initial-A and the weak-coupling
method correspond to the same physical initial state. The delay observed
only in \ensuremath{\mid}A\textminus weak\ensuremath{\mid} therefore
suggests that the variational polaron transformation of the initial
state in Initial-A is properly incorporated into the corresponding
master equation.

\begin{figure}
\includegraphics[width=1\columnwidth]{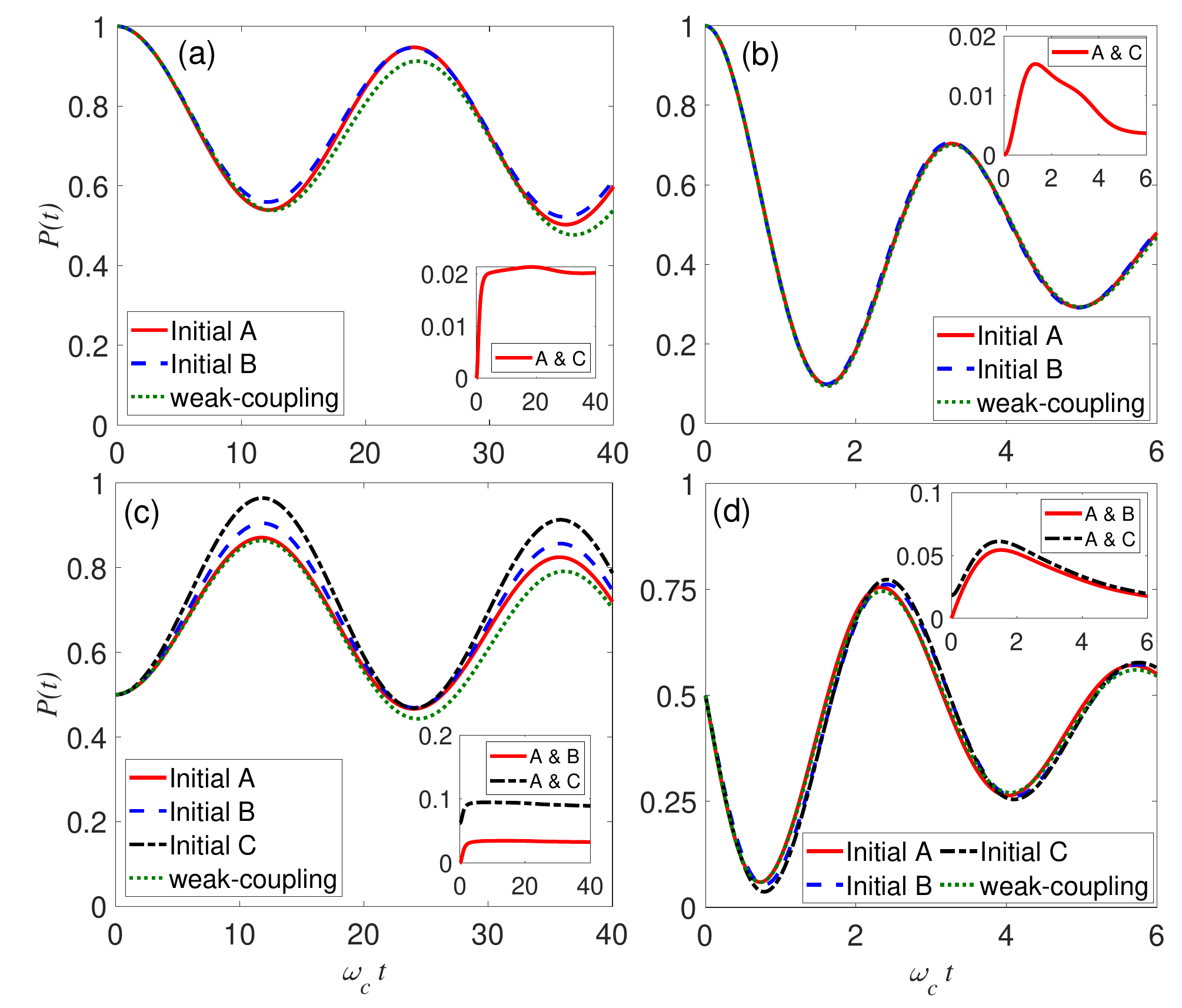}
\caption{Dynamics of the population $P\left(t\right)$ and the corresponding
time evolution of the trace distances at zero temperature for the
setups considered in Fig. \ref{fig:P(t)_spin-boson_T_4}. The curves
in the main panels and insets follow the same notation as in Fig.
\ref{fig:P(t)_spin-boson_T_4}. The initial states and other parameters
in panels (a)--(d) are the same as those in Figs. \ref{fig:P(t)_spin-boson_T_4}(a)--\ref{fig:P(t)_spin-boson_T_4}(d),
respectively, except that $T=0$.}
\label{fig:P(t)_zero_T}
\end{figure}

\subsubsection{Dependence on temperature}

Having discussed the dependence of the system dynamics on the qubit
tunneling amplitude $\Delta$, we next investigate the temperature
dependence by considering $T=0$ as the lower-temperature counterpart
of the cases shown in Fig. \ref{fig:P(t)_spin-boson_T_4}, with the
corresponding results presented in Fig. \ref{fig:P(t)_zero_T}. At
lower temperature, the predictions of the weak-coupling and variational
polaron methods become closer for the parameters considered. In particular,
the population dynamics obtained from the weak-coupling method (green
dotted line) approach those of the variational polaron method in Figs.
\ref{fig:P(t)_zero_T}(a) and \ref{fig:P(t)_zero_T}(c), in contrast
to the breakdown of the weak-coupling results observed in Figs. \ref{fig:P(t)_spin-boson_T_4}(a)
and \ref{fig:P(t)_spin-boson_T_4}(c) at higher temperature.

A comparison of the trace distances in the insets of Fig. \ref{fig:P(t)_zero_T}
with their higher-temperature counterparts in Fig. \ref{fig:P(t)_spin-boson_T_4}
reveals a more gradual decay at lower temperature, indicating a longer
persistence of the initial-state-induced effects and being consistent
with slower bath relaxation. For the smaller tunneling amplitude $\Delta=0.2\omega_{c}$
(left panels), the decay of the trace distances becomes particularly
gradual at lower temperature.

For the initial state $|+\rangle_{z}$, Initial-B and Initial-C are
identical. The corresponding trace distance in the inset of Fig. \ref{fig:P(t)_spin-boson_T_4}(a)
is approximately an order of magnitude smaller than that in the inset
of Fig. \ref{fig:P(t)_zero_T}(a), indicating a smaller contribution
of the inhomogeneous terms at higher temperature. In this case, the
relevant functions arise from the diagonal matrix elements , with
$C_{1}^{\textrm{d}_{1}}\left(t\right)$ and $C_{2}^{\textrm{d}_{1}}\left(t\right)$
being the dominant contributions {[}see Eqs. (\ref{eq:C_1^d_1(t)})
and (\ref{eq:C_2^d_1(t)}){]}. These functions contain the proportional
factor $\eta$, which decreases monotonically as the temperature increases
{[}see Eq. (\ref{eta(w)}){]}. At sufficiently high temperature, $\eta\rightarrow0$,
and consequently $C_{i}^{\textrm{d}_{1}}\left(t\right)\rightarrow0$.
This suppression, together with the faster decay of the trace distance
observed at higher temperature, indicates a separation of timescales
in which the bath relaxes much faster than the system dynamics. Under
such conditions, the distinction between taking the initial bath as
a displaced bath or as a thermal equilibrium bath rapidly becomes
insignificant. The corresponding inhomogeneous contributions therefore
become negligible, and the system dynamics are predominantly governed
by the homogeneous part of the variational polaron master equation.

The situation is different for the initial states $|+\rangle_{x}$
and $|-\rangle_{y}$, for which contributions associated with both
the diagonal and off-diagonal matrix elements are involved. For the
off-diagonal contributions, the factor $\exp\left(-\textrm{Re}\left\{ r\left(0\right)\right\} \right)=\eta^{-2}$
{[}see $r\left(t\right)$ in Eq. (\ref{r(tau)}){]} appearing in the
dominant functions $C_{1}^{\textrm{d}_{2}}\left(t\right)$ and $C_{2}^{\textrm{d}_{2}}\left(t\right)$
{[}see Eqs. (\ref{eq:C_1^d_2(t)}) and (\ref{eq:C_2^d_2(t)}){]} can
compensate for the decrease of the proportional factor $\eta$ as
the temperature increases. Consequently, the inhomogeneous contributions
associated with the off-diagonal matrix elements do not necessarily
diminish and may even be enhanced at higher temperature. This behavior
is reflected in the higher peaks of the trace distance between \textquotedblleft A
\& C\textquotedblright{} in the insets of Figs. \ref{fig:P(t)_spin-boson_T_4}(c)
and \ref{fig:P(t)_spin-boson_T_4}(d) compared with those in Figs.
\ref{fig:P(t)_zero_T}(c) and \ref{fig:P(t)_zero_T}(d). In this case,
properly accounting for the transformed initial state and the associated
inhomogeneous terms is therefore necessary for a full description
of the dynamics.

\subsection{Dynamics of the expectation values $\left\langle \sigma_{x}\left(t\right)\right\rangle $
and $\left\langle \sigma_{y}\left(t\right)\right\rangle $}

\label{sec:Off_diagonal_expection_value}

\begin{figure}
\includegraphics[width=1\columnwidth]{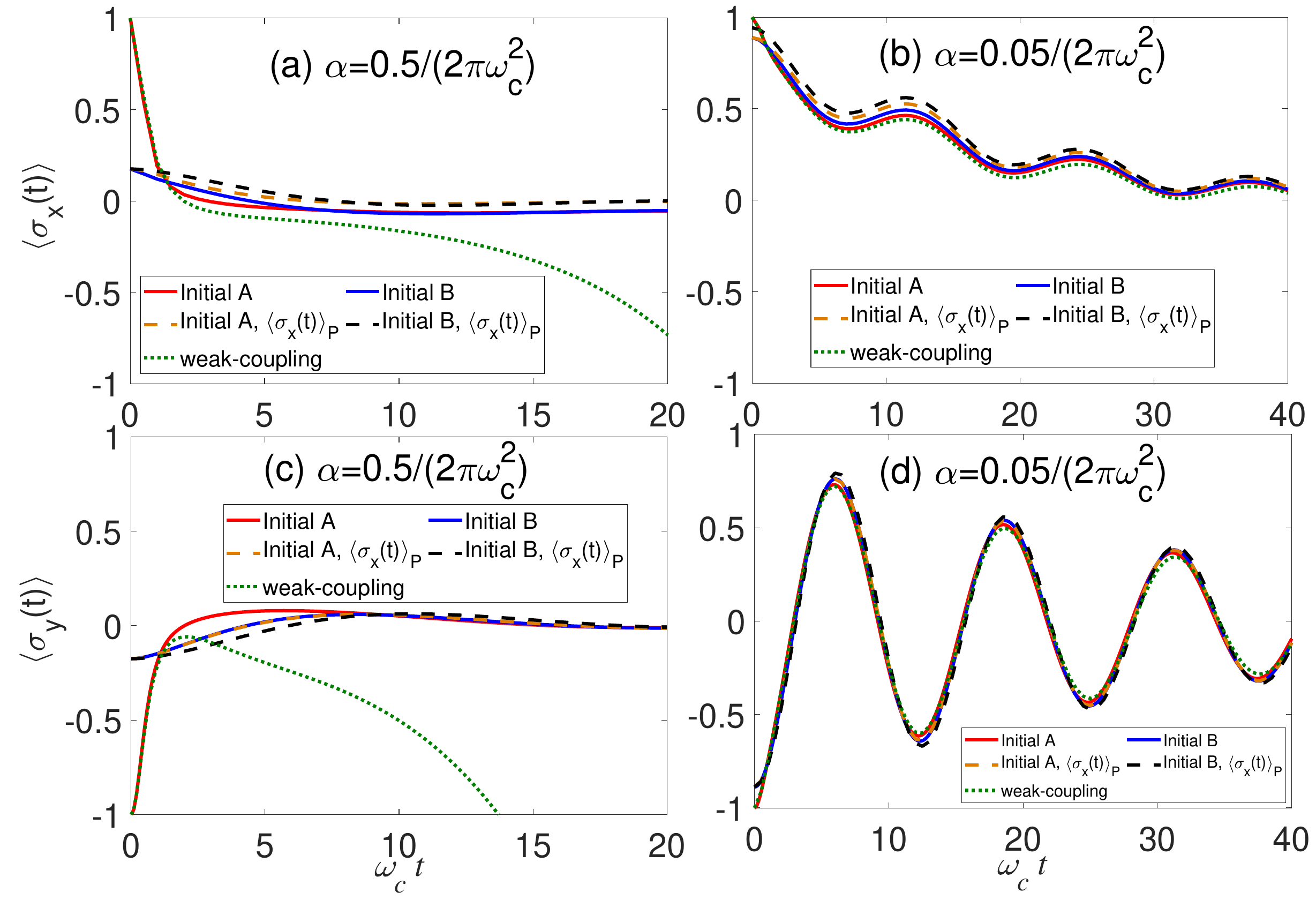}\caption{Dynamics of $\left\langle \sigma_{x}\left(t\right)\right\rangle $
in panels (a) and (b) and $\left\langle \sigma_{y}\left(t\right)\right\rangle $
in panels (c) and (d), calculated using the variational polaron method
with Initial-A (red) and Initial-B (blue), together with the weak-coupling
method (dark-green dotted line). The initial states and system-bath
coupling strengths are: (a) $|+\rangle_{x}$, $\alpha=0.5/\left(2\pi\omega_{c}^{2}\right)$;
(b) $|+\rangle_{x}$, $\alpha=0.05/\left(2\pi\omega_{c}^{2}\right)$;
(c) $|-\rangle_{y}$, $\alpha=0.5/\left(2\pi\omega_{c}^{2}\right)$;
and (d) $|-\rangle_{y}$, $\alpha=0.05/\left(2\pi\omega_{c}^{2}\right)$.
For the variational polaron method, the solid lines show the full
expectation values obtained from Eq. (\ref{eq:sigma_P_Q}), whereas
the dashed lines show the corresponding $P$-projected parts, $\left\langle \sigma_{x}\left(t\right)\right\rangle _{P}$
and $\left\langle \sigma_{y}\left(t\right)\right\rangle _{P}$. Other
parameters are $\Delta=0.5\omega_{c}$, $\varepsilon=0.2\omega_{c}$,
and $k_{B}T=4\omega_{c}$.}
\label{fig:coherence_dynamics} 
\end{figure}

After solving Eq. (\ref{VP master eq}), the dynamics of $\left\langle \sigma_{-}\left(t\right)\right\rangle $
can be obtained directly from Eq. (\ref{eq:sigma_P_Q}) using the
previously calculated $\rho_{s}'\left(t\right)$ and the auxiliary
matrices given in Eqs. (\ref{dot_D_ij,m(t)})-(\ref{dot_D_^dq_i(t)}).
Thus, no separate dynamical equation needs to be propagated to obtain
$\left\langle \sigma_{x}\left(t\right)\right\rangle =2\textrm{Re}\left\{ \left\langle \sigma_{-}\left(t\right)\right\rangle \right\} $
and $\left\langle \sigma_{y}\left(t\right)\right\rangle =-2\textrm{Im}\left\{ \left\langle \sigma_{-}\left(t\right)\right\rangle \right\} $.
Importantly, Eq. (\ref{eq:sigma_P_Q}) also incorporates the effects
associated with the transformed initial states.

Having explored the regimes of large and small $\Delta$ in Fig. \textcolor{blue}{\ref{fig:P(t)_spin-boson_T_4}},
here we choose an intermediate value $\Delta=0.5\omega_{c}$ and consider
both large and small values of the system-bath coupling constant $\alpha$.
The results obtained from the full expression in Eq. (\ref{eq:sigma_P_Q})
are shown by the solid lines in Fig. \ref{fig:coherence_dynamics}.
For comparison, we also consider a simplified treatment retaining
only the corresponding $P$-projected contribution (dashed lines),
defined by $\left\langle \sigma_{x}\left(t\right)\right\rangle _{P}=2\textrm{Re}\left\{ \left\langle \sigma_{-}\left(t\right)\right\rangle _{P}\right\} $
and $\left\langle \sigma_{y}\left(t\right)\right\rangle _{P}=-2\textrm{Im}\left\{ \left\langle \sigma_{-}\left(t\right)\right\rangle _{P}\right\} $.
By construction, at $t=0$, $\left\langle \sigma_{-}\left(0\right)\right\rangle _{P}=\eta\textrm{tr}_{s}\left[\sigma_{-}\rho_{s}'\left(0\right)\right]=\eta^{2}\rho_{10}$
for both Initial-A and Initial-B, and therefore has the same initial
value as the full $\left\langle \sigma_{-}\left(t\right)\right\rangle $
obtained from Initial-B. Consequently, either employing Initial-B
or retaining only the $P$-projected contribution leads to a deviated
initial expectation value and subsequently modified dynamics. As shown
in Fig. \ref{fig:coherence_dynamics}(a), where $\left\langle \sigma_{x}\left(t\right)\right\rangle $
approaches a nonzero long-time equilibrium value, $\left\langle \sigma_{x}\left(t\right)\right\rangle _{P}$
instead approaches zero and therefore deviates from the full result
at long times. In contrast, in Fig. \ref{fig:coherence_dynamics}(c),
where the equilibrium value of $\left\langle \sigma_{y}\left(t\right)\right\rangle $
approaches zero, $\left\langle \sigma_{y}\left(t\right)\right\rangle _{P}$
approaches the full result in the long-time limit.

For comparison, the results obtained from the weak-coupling method
are shown by the dark-green dotted lines in Fig. \ref{fig:coherence_dynamics}.
In Figs. \ref{fig:coherence_dynamics}(a) and \ref{fig:coherence_dynamics}(c),
the results from the variational polaron method with Initial-A exhibit
an initial overlap with those from the weak-coupling method over the
time scale $t\sim\omega_{c}^{-1}$, analogous to the delay behavior
discussed in Fig. \ref{fig:multi-bath-quanta_delay}. Figures \ref{fig:coherence_dynamics}(b)
and \ref{fig:coherence_dynamics}(d) show the corresponding results
for the smaller system-bath coupling constant $\alpha=0.05/\left(2\pi\omega_{c}^{2}\right)$,
where the system lies in the weak-coupling regime. In this regime,
the results of the variational polaron method, particularly those
obtained with Initial-A (red solid lines), become significantly closer
to those of the weak-coupling method (dark-green dotted lines). The
short-time overlap and the convergence toward the weak-coupling results
as the coupling strength decreases further support the use of Eq.
(\ref{eq:sigma_P_Q}) together with Initial-A for evaluating the off-diagonal
expectation values.

Finally, we summarize the effects of the initial states on the dynamics
in Table \ref{table_Effect_initial_state}. The assessment summarized
in Table II is made by comparison with the numerically exact HEOM
dynamics, with which the Initial-A results show close agreement. The
initial state $|+\rangle_{z}$ exhibits a relatively small initial-state
effect on the population dynamics. In contrast, more pronounced effects
are observed in $\left\langle \sigma_{x}\left(t\right)\right\rangle $
and $\left\langle \sigma_{y}\left(t\right)\right\rangle $ when both
the initial-state density operator, corresponding to $|+\rangle_{x}$
or $|-\rangle_{y}$, and the observable do not commute with the variational
polaron transformation.

\begin{table*}
\begin{tabular}{|c|c|c|}
\hline 
Lab-frame initial state & $|+\rangle_{z}$ & $|+\rangle_{x}$ or $|-\rangle_{y}$\tabularnewline
\hline 
Treatment & Initial-A (= Initial-B)/Initial-C & Initial-A/Initial-B/Initial-C\tabularnewline
\hline 
$\left\langle \sigma_{z}\left(t\right)\right\rangle $ & $\surd/\bigcirc$ & $\surd/\diamondsuit/\diamondsuit$\tabularnewline
\hline 
$\left\langle \sigma_{x}\left(t\right)\right\rangle $ and $\left\langle \sigma_{y}\left(t\right)\right\rangle $ & $\surd/\diamondsuit$ & $\surd/\oslash/\oslash$\tabularnewline
\hline 
$\left\langle \sigma_{x}\left(t\right)\right\rangle _{P}$ & $\widetilde{\diamondsuit}/\widetilde{\diamondsuit}$ & $\widetilde{\oslash}/\widetilde{\oslash}/\widetilde{\oslash}$\tabularnewline
\hline 
$\left\langle \sigma_{y}\left(t\right)\right\rangle _{P}$ & $\diamondsuit/\diamondsuit$ & $\oslash/\oslash/\oslash$\tabularnewline
\hline 
\multicolumn{3}{|l|}{$\surd$: correct initial value and dynamics,}\tabularnewline
\hline 
\multicolumn{3}{|l|}{$\bigcirc$: small effect on the dynamics,}\tabularnewline
\hline 
\multicolumn{3}{|l|}{$\diamondsuit$: affects the dynamics,}\tabularnewline
\hline 
\multicolumn{3}{|l|}{$\oslash$: affects both the initial value and the dynamics,}\tabularnewline
\hline 
\multicolumn{3}{|l|}{$\widetilde{\;\;}$: the effect persists at long times.}\tabularnewline
\hline 
\end{tabular}

\caption{Effects of the polaron-transformed initial state (Initial-A) and the
two simplified initial-state treatments (Initial-B and Initial-C)
on the dynamics of the expectation values for the laboratory-frame
initial states $|+\rangle_{z}$, $|+\rangle_{x}$ and $|-\rangle_{y}$.
Symbols separated by slashes correspond, from left to right, to the
initial-state treatments listed in the second row. A tilde over a
symbol indicates that the corresponding effect persists at long times.
Here, $\left\langle \sigma_{x}\left(t\right)\right\rangle _{P}$ and
$\left\langle \sigma_{y}\left(t\right)\right\rangle _{P}$ denote
the $P$-projected parts of $\left\langle \sigma_{x}\left(t\right)\right\rangle $
and $\left\langle \sigma_{y}\left(t\right)\right\rangle $, respectively.
For the initial states $|+\rangle_{x}$ and $|-\rangle_{y}$, the
deviations between $\left\langle \sigma_{y}\left(t\right)\right\rangle _{P}$
and $\left\langle \sigma_{y}\left(t\right)\right\rangle $ vanish
at long times for all three initial-state treatments, possibly because
$\left\langle \sigma_{y}\left(t\right)\right\rangle $ approaches
zero within the parameter regime considered.}

\label{table_Effect_initial_state}
\end{table*}

\section{Driven spin-boson model}

\label{sec:driven_spin_boson}

Previous work \citep{A_general_RWA_VP_master_equation} employing
the variational polaron approach for a driven system focused primarily
on the population dynamics of a continuously driven quantum dot (QD).
In that model, the bath couples only to the QD excited state, such
that initialization of the QD in its ground state remains unaffected
by the variational polaron transformation. Thus, both the population
observable and the chosen initial state are invariant under the transformation.
To extend the variational polaron approach to driven systems in which
both the initial state and observables are transformed, we consider
a driven spin-boson model and apply our method to this case.

In contrast to the time-independent case in Eq. (\ref{H_T}), the
coefficient of $\sigma_{x}$ in the system Hamiltonian is now given
by the time-dependent driving field $\Omega\left(t\right)=\Omega_{0}\cos\left(\omega_{L}t\right)$,
where $\omega_{L}$ and $\Omega_{0}$ are the frequency and amplitude
of the externally applied driving field, respectively. Accordingly,
the coefficient $\Delta/2$ in Eq.~(\ref{H_s}) and Eq.~(\ref{H'_s})
is replaced by $\Omega\left(t\right)$. Following Eqs.~(\ref{H'_T})-(\ref{H'_sb}),
the corresponding Hamiltonian in the variational polaron frame $\left(\hbar=1\right)$
is
\begin{eqnarray}
H_{T}'\left(t\right) & = & H_{s}'\left(t\right)+H_{sb}'\left(t\right)+H_{b},\label{H'_T(t)}\\
H_{s}'\left(t\right) & = & \frac{\epsilon}{2}\sigma_{z}+\eta\Omega\left(t\right)\sigma_{x}+R\sigma_{0},\label{H'_s(t)}\\
H_{sb}'\left(t\right) & = & \Omega\left(t\right)\left(B_{1}\sigma_{x}+B_{2}\sigma_{y}\right)+B_{3}\sigma_{z},\label{H'_sb(t)}
\end{eqnarray}
where $B_{i}$ with $i=1,2,3,$ are defined in Eqs.~(\ref{eq:B_1})-(\ref{eq:B_3}). 

To perform the free-energy minimization, Eq.~(\ref{H'_T(t)}) is
transformed into the rotating frame. The total density state in the
polaron-rotating frame is defined as $\rho_{T}^{r'}\left(t\right)=e^{S_{L}\left(t\right)}\rho_{T}'\left(t\right)e^{-S_{L}\left(t\right)}$,
where $S_{L}\left(t\right)=i\frac{\omega_{L}}{2}\sigma_{z}t$. From
the equation of motion $\frac{d}{dt}\rho_{T}'\left(t\right)=-\frac{i}{\hbar}\left[H_{T}'\left(t\right),\rho_{T}'\left(t\right)\right]$
in the polaron frame, it follows that
\begin{eqnarray}
\frac{d}{dt}\rho_{T}^{r'}\left(t\right) & = & -\frac{i}{\hbar}\left[H_{T}^{r'}\left(t\right),\rho_{T}^{r'}\left(t\right)\right],
\end{eqnarray}
with the effective Hamiltonian
\begin{eqnarray}
H_{T}^{r'}\left(t\right) & = & e^{S_{L}\left(t\right)}H_{T}'\left(t\right)e^{-S_{L}\left(t\right)}+i\frac{d}{dt}S_{L}\left(t\right)\nonumber \\
 & = & H_{s}^{r'}\left(t\right)+H_{sb}^{r'}\left(t\right)+H_{b},\\
H_{s}^{r'}\left(t\right) & = & \frac{\delta}{2}\sigma_{z}+\frac{\eta\Omega_{0}}{2}\sigma_{x}+R\sigma_{0}\nonumber \\
 &  & +\frac{\eta\Omega_{0}}{2}\left(\sigma_{+}e^{2i\omega_{L}t}+\sigma_{-}e^{-2i\omega_{L}t}\right),\label{eq:CR_term_2wL_Hs}\\
H_{sb}^{r'}\left(t\right) & = & B_{3}\sigma_{z}+B_{1}\frac{\Omega_{0}}{2}\left(\sigma_{x}+\sigma_{+}e^{2i\omega_{L}t}+\sigma_{-}e^{-2i\omega_{L}t}\right)\nonumber \\
 &  & +B_{2}\frac{\Omega_{0}}{2}\left[\sigma_{y}-i\left(\sigma_{+}e^{2i\omega_{L}t}-\sigma_{-}e^{-2i\omega_{L}t}\right)\right].\label{eq:CR_term_2wL_Hsb}
\end{eqnarray}
Subsequently, the rotating-wave approximation (RWA) is applied by
neglecting the rapidly oscillating terms with frequencies $\pm2\omega_{L}$
in Eqs.~(\ref{eq:CR_term_2wL_Hs}) and (\ref{eq:CR_term_2wL_Hsb}).
This yields the time-independent Hamiltonian in the polaron-rotating
frame, $H_{T}^{r'}=H_{s}^{r'}+H_{sb}^{r'}+H_{b}$ where
\begin{eqnarray}
H_{s}^{r'} & = & \frac{\delta}{2}\sigma_{z}+\frac{\eta\Omega_{0}}{2}\sigma_{x}+R\sigma_{0},\\
H_{sb}^{r'} & = & \frac{\Omega_{0}}{2}\left(B_{1}\sigma_{x}+B_{2}\sigma_{y}\right)+B_{3}\sigma_{z}.
\end{eqnarray}
Here, $\delta=\epsilon-\omega_{L}$ is the detuning. Similar to the
time-independent case \citep{A_general_RWA_VP_master_equation}, the
variational parameter $\xi_{k}$ is determined as
\begin{equation}
\xi_{k}=\left[1+\frac{X_{\textrm{RWA}}}{\omega_{k}}\coth\left(\frac{\beta\omega_{k}}{2}\right)\tanh\left(\beta W_{\textrm{RWA}}\right)\right]^{-1},\label{xi_k_RWA}
\end{equation}
where $W_{\textrm{RWA}}=\sqrt{\left(\eta\Omega_{0}/2\right)^{2}+\left(\delta/2\right)^{2}}$
and $X_{\textrm{RWA}}=\left(\eta\Omega_{0}\right)^{2}/\left(2W_{\textrm{RWA}}\right)$.
The derivation of the corresponding master equation for $\rho_{s}^{r'}\left(t\right)=\textrm{tr}_{b}\left[\rho_{T}^{r'}\left(t\right)\right]$,
governed by $H_{T}^{r'}$, closely follows that for the time-independent
case presented in Sec. \ref{sec:Variational-polaron_transformation},
with the replacements $\varepsilon\rightarrow\delta$ and $\Delta\rightarrow\Omega_{0}$.
The expectation values $\left\langle \sigma_{z}\left(t\right)\right\rangle =\textrm{tr}_{s}\left[\sigma_{z}\rho_{s}^{r'}\left(t\right)\right]$
and $\left\langle \sigma_{-}\left(t\right)\right\rangle =\textrm{tr}_{T}\left[\sigma_{-}'e^{-i\omega_{L}t}\rho_{T}^{r'}\left(t\right)\right]$
can then be evaluated following the procedure described in Sec.~\ref{sec:Off_diagonal_expection_value},
with the latter containing the additional phase factor $\exp\left(-i\omega_{L}t\right)$.

\begin{figure}
\includegraphics[width=1\columnwidth]{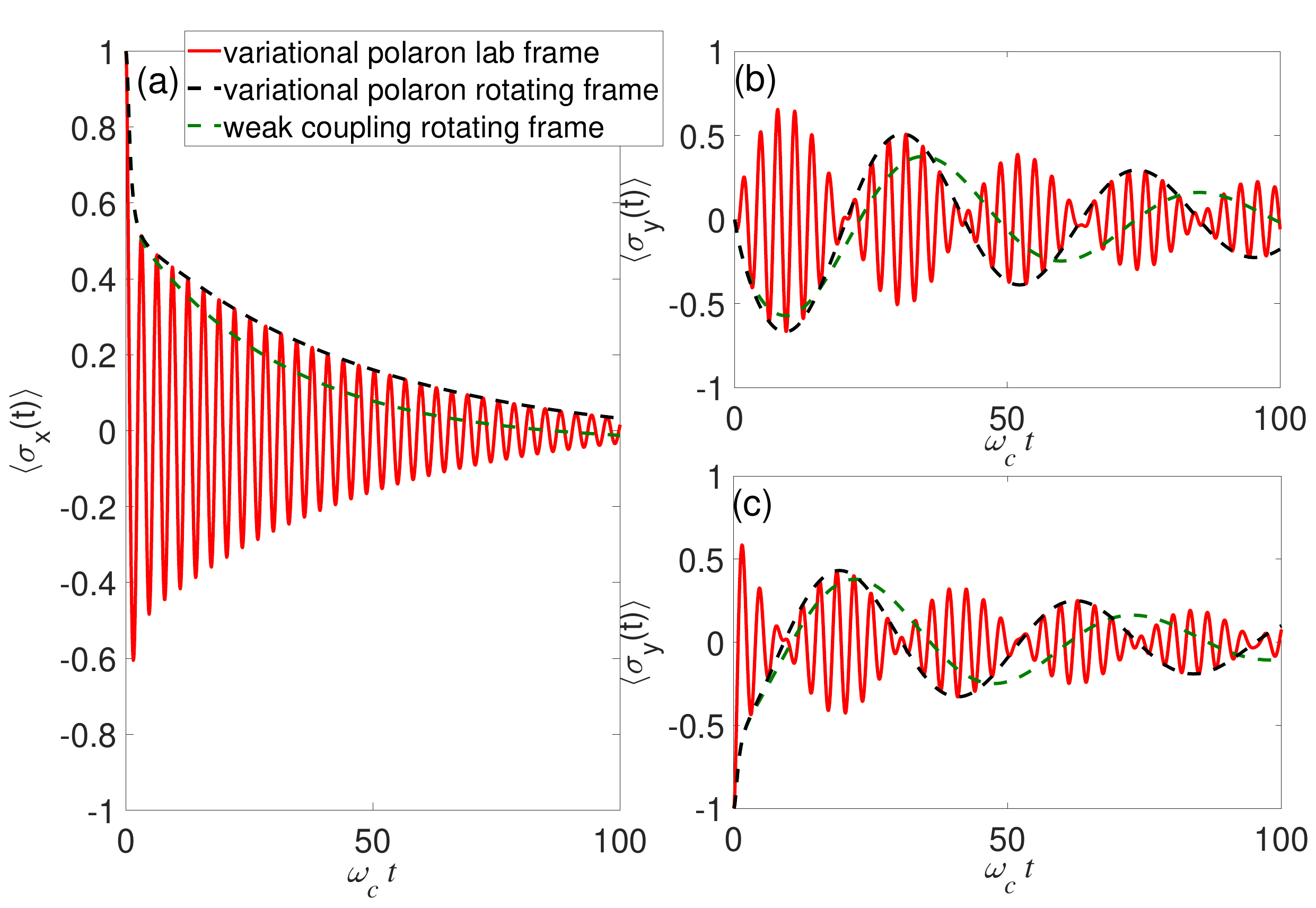}\caption{Dynamics of $\left\langle \sigma_{x}\left(t\right)\right\rangle $
and $\left\langle \sigma_{y}\left(t\right)\right\rangle $ for the
resonantly driven spin-boson model with the driving term $\Omega_{0}\cos\left(\omega_{L}t\right)\sigma_{x}$.
The red solid lines show the laboratory-frame results obtained using
the variational polaron method with Initial-A, while the red dashed
lines show the corresponding dynamics in the rotating frame. The green
dashed lines show the rotating-frame results obtained using the weak-coupling
method. The corresponding initial states in the laboratory frame are
(a) $|+\rangle_{x}$, (b) $|+\rangle_{z}$, and (c) $|-\rangle_{y}$.
Other parameters are $\alpha=0.17/\left(2\pi\omega_{c}^{2}\right)$,
$\Omega_{0}=0.2\omega_{c}$, $\omega_{L}=\varepsilon=2\omega_{c}$,
and $k_{B}T=3\omega_{c}$.}
\label{fig:driven_coherence_dynamics} 
\end{figure}

Figure \ref{fig:driven_coherence_dynamics} presents the dynamics
of $\left\langle \sigma_{x}\left(t\right)\right\rangle $ and $\left\langle \sigma_{y}\left(t\right)\right\rangle $
for resonant driving, with the $Q$-projected contributions and the
transformed initial state (Initial-A) fully included. In the laboratory
frame, the dynamics exhibit rapid oscillations at $\omega_{L}=2\omega_{c}$,
as shown by the red solid lines. Removing the phase factor $\exp\left(-i\omega_{L}t\right)$
gives $\left\langle \sigma_{-}\left(t\right)\right\rangle ^{r}=\textrm{tr}_{T}\left[\sigma_{-}'\rho_{T}^{r'}\left(t\right)\right]$,
which corresponds to an effective measurement in the rotating frame,
as shown by the black dashed lines. In the resonant, weak-driving
regime considered here, with relatively strong system-bath coupling
and elevated temperature, the weak-coupling results (green dashed
lines) deviate significantly from those obtained using the variational
polaron approach in the rotating frame, indicating that these parameters
lie beyond the regime of validity of the weak-coupling approximation.
A related method for effectively performing measurements in the rotating
frame has been experimentally demonstrated in Ref. \citep{NV_center_measure_in_rotatng_frame},
where the phase of a sequential pulse is calibrated according to $\exp\left(-i\omega_{L}t\right)$,
thereby removing oscillations at the driving frequency $\omega_{L}$.

In Fig. \ref{fig:driven_coherence_dynamics}(a), the initial state
$|+\rangle_{x}$ is an eigenstate of the driving Hamiltonian, and
consequently $\left\langle \sigma_{x}\left(t\right)\right\rangle $
exhibits a monotonic decay in the rotating frame. By contrast, for
initial states that are not eigenstates of the driving Hamiltonian,
as shown in Figs. \ref{fig:driven_coherence_dynamics}(b) and \ref{fig:driven_coherence_dynamics}(c),
the dynamics of $\left\langle \sigma_{y}\left(t\right)\right\rangle $
exhibit different decoherence patterns, accompanied by slower oscillations
at the driving amplitude $\Omega_{0}=0.2\omega_{c}$ in both frames.
These results show that the rapid oscillations at $\omega_{L}$ appearing
in the laboratory frame are removed in the rotating frame and demonstrate
that the method developed in Sec. \ref{sec:Off_diagonal_expection_value}
can be readily applied to driven systems within the variational polaron
master-equation approach while consistently incorporating the transformed
initial state.

For an independent numerical validation, all Initial-A results presented
in Secs. \ref{sec:Dynamics_result_undriven} and \ref{sec:driven_spin_boson}
have been benchmarked against numerically exact hierarchical equations
of motion (HEOM) calculations \citep{HEOM_Huang2023,QuantumToolbox.jl2025}
and found to be in good agreement, with four representative comparisons
shown in Appendix \ref{sec:appendix_HEOM_benchmark}.

\section{Conclusion}

\label{sec:Conclusion}

We have developed a variational polaron master equation framework
that consistently incorporates the effects of the transformed initial
state on the dynamics of the spin-boson model. Starting from a factorized
initial state in the laboratory frame, the polaron transformation
generally produces a correlated system-bath state containing both
$P$- and $Q$-projected components. The latter gives rise to the
inhomogeneous terms in the variational polaron master equation and
therefore accounts for the effects of the transformed initial state
that are neglected when only the $P$-projected component is retained.
By reformulating the resulting integro-differential equations in an
extended auxiliary Liouville space, the dynamics can be efficiently
evaluated through a set of coupled differential equations.

We have also addressed the transformation of observables that do not
commute with the variational polaron transformation. In particular,
we have introduced a systematic method for calculating the dynamics
of $\left\langle \sigma_{x}\left(t\right)\right\rangle $ and $\left\langle \sigma_{y}\left(t\right)\right\rangle $,
while consistently incorporating the effects of the transformed initial
state. Together with the directly accessible population $P\left(t\right)=[1+\langle\sigma_{z}\left(t\right)\rangle]/2$,
this treatment provides the dynamics of the reduced system in the
original laboratory frame without restricting the initial system state
or the observable under consideration. In the pure-dephasing limit,
our formulation recovers the exact analytical result, providing a
consistency check on the treatment of both the transformed initial
state and observables.

Our numerical results demonstrate that the transformation of the initial
state can have a pronounced influence on the system dynamics, particularly
at short times. The effects are generally stronger for the coherent
superposition states $|+\rangle_{x}$ and $|-\rangle_{y}$ than for
the localized state $|+\rangle_{z}$, and their relative importance
depends on the qubit tunneling amplitude $\Delta$, temperature $T$,
and system-bath coupling strength $\alpha$. Comparisons among the
complete transformed initial state (Initial-A) and the simplified
Initial-B and Initial-C states, together with the corresponding trace
distances, clarify the distinct roles of the $P$- and $Q$-projected
components of the transformed initial state. In addition, comparison
with the weak-coupling master equation reveals a delay in the deviation
of the population dynamics obtained from the two approaches. The characteristic
delay time exhibits an approximately inverse dependence on $\Delta$
over the investigated intermediate regime, suggesting a distinct timescale
associated with higher-order multi-bath-quanta processes captured
by the variational polaron approach.

We have further extended the treatment of the transformed initial
state and observables to a resonantly driven spin-boson model. The
dynamics evaluated in both the rotating and laboratory frames demonstrate
that the effects of the transformed initial state depend on the chosen
initial state and observable and remain relevant in the presence of
an external driving field. All Initial-A results presented in this
work have also been verified against numerically exact hierarchical
equations of motion (HEOM) calculations, demonstrating the reliability
of the present approach in the parameter regimes considered.

Overall, the present formulation provides a versatile and computationally
efficient variational polaron master equation approach for treating
open quantum systems with general factorized initial states in the
laboratory frame and observables that may or may not commute with
the polaron transformation. By consistently accounting for the transformed
initial state and observables, the framework extends the applicability
of variational polaron methods across a broad range of system-bath
coupling strengths and parameter regimes. A natural extension of the
present work is to generalize the treatment of the driven spin-boson
model beyond the rotating-wave approximation within the variational
polaron framework.
\begin{acknowledgments}
The authors acknowledge fruitful discussions with Yi-Te Huang, the
developer of the HierarchicalEOM.jl package, regarding the implementation
of the HEOM calculations. H.-S. Goan acknowledges the support from
NSTC, Taiwan, under Grants No.~NSTC 113-2112-M-002-022-MY3, No.~NSTC
113-2119M-002-021, No.~NSTC 114-2119-M-002-018, No.~NSTC 114-2119-M-002-017-MY3,
and from the National Taiwan University under Grant No.~NTU-CC-115L8937,
No.~NTU-CC-115L893704 and No.~NTU-CC-115L8512. H.-S. Goan is also
grateful for the support of the "Center for Advanced Computing and
Imaging in Biomedicine (NTU-115L900702)" through the Featured Areas
Research Center Program within the framework of the Higher Education
Sprout Project by the Ministry of Education (MOE), Taiwan, the support
of Taiwan Semiconductor Research Institute (TSRI) through the Joint
Developed Project (JDP) and the support from the Physics Division,
National Center for Theoretical Sciences, Taiwan. 
\end{acknowledgments}

\subsection*{AUTHOR CONTRIBUTIONS}

C.-C.C. developed the theoretical formulation, performed the numerical
calculations, analyzed the results, and prepared the manuscript. H.-S.G.
supervised the research, contributed to the analysis and discussion,
and reviewed and edited the manuscript. Both authors approved the
final manuscript. 
\appendix

\section{Inhomogeneous terms of the variational polaron master equation}

\label{sec:appendixA}

In this appendix, we present the details of the functions and matrices
that constitute the inhomogeneous terms of the variational polaron
master equation arising from the transformed initial states. Upon
expansion of $Q\tilde{\rho}_{T}'\left(t\right)$ in Eq. \eqref{d_dt P rho_T(t) inter pic}
to first order in $\tilde{H}_{sb}'$, the first-order inhomogeneous
term associated with the $Q$-projected part of the transformed initial
state is given by \citep{Breuer02,VP_Master_Eq_DC_and_Imhomo}
\begin{equation}
\textrm{tr}_{b}\left[\tilde{\mathscr{L}}_{sb}'\left(t\right)Q\tilde{\rho}_{T}'\left(0\right)\right]=-i\textrm{tr}_{b}\left[\tilde{H}_{sb}'\left(t\right),Q\tilde{\rho}_{T}'\left(0\right)\right],\label{eq:Q1}
\end{equation}
and the corresponding second-order term is
\begin{eqnarray}
 &  & \int_{0}^{t}\textrm{tr}_{b}\left[\tilde{\mathscr{L}}_{sb}'\left(t\right)\tilde{\mathscr{L}}_{sb}'\left(t'\right)Q\tilde{\rho}_{T}'\left(0\right)\right]dt'\nonumber \\
 & = & -\int_{0}^{t}\textrm{tr}_{b}\left[\tilde{H}_{sb}'\left(t\right),\left[\tilde{H}_{sb}'\left(t'\right),Q\tilde{\rho}_{T}'\left(0\right)\right]\right]dt'.\label{eq:Q2}
\end{eqnarray}
After performing the trace in Eq. \eqref{eq:Q1} over the bath degrees
of freedom, we obtain the first-order functions $C_{i}^{\textrm{d},\left(u\right)}\left(t\right)=\textrm{tr}_{b}\left[e^{iH_{b}t}B_{i}e^{-iH_{b}t}\rho_{b,\left(u\right)}^{\textrm{d}}\right]$.
For the diagonal components ($q=1$),
\begin{align}
C_{1}^{\textrm{d}_{1}}\left(t\right) & =C_{1}^{\textrm{d},\left(1\right)}\left(t\right)=C_{1}^{\textrm{d},\left(4\right)}\left(t\right)=\eta\left[\cos\left(\textrm{Im}\left\{ r\left(t\right)\right\} \right)-1\right],\label{eq:C_1^d_1(t)}\\
C_{2}^{\textrm{d}_{1}}\left(t\right) & =C_{2}^{\textrm{d},\left(1\right)}\left(t\right)=-C_{2}^{\textrm{d},\left(4\right)}\left(t\right)=-\eta\sin\left(\textrm{Im}\left\{ r\left(t\right)\right\} \right),\label{eq:C_2^d_1(t)}\\
C_{3}^{\textrm{d}_{1}}\left(t\right) & =C_{3}^{\textrm{d},\left(1\right)}\left(t\right)=-C_{3}^{\textrm{d},\left(4\right)}\left(t\right)=\eta^{-1}\textrm{Im}\left[C_{32}\left(t\right)\right].
\end{align}
For the off-diagonal components ($q=2$),
\begin{align}
C_{1}^{\textrm{d}_{2}}\left(t\right) & =C_{1}^{\textrm{d},\left(2\right)}\left(t\right)=C_{1}^{\textrm{d},\left(3\right)}\left(t\right)=\eta\left[\cosh\left(\textrm{Re}\left\{ r\left(t\right)\right\} \right)-1\right],\label{eq:C_1^d_2(t)}\\
C_{2}^{\textrm{d}_{2}}\left(t\right) & =C_{2}^{\textrm{d},\left(2\right)}\left(t\right)=-C_{2}^{\textrm{d},\left(3\right)}\left(t\right)=-i\eta\sinh\left(\textrm{Re}\left\{ r\left(t\right)\right\} \right),\label{eq:C_2^d_2(t)}\\
C_{3}^{\textrm{d}_{2}}\left(t\right) & =C_{3}^{\textrm{d},\left(2\right)}\left(t\right)=-C_{3}^{\textrm{d},\left(3\right)}\left(t\right)=i\eta^{-1}\textrm{Re}\left\{ C_{32}\left(t\right)\right\} .\label{eq:C_3^d_2(t)}
\end{align}
Here,
\[
r\left(t\right)=\ln\left[C_{xx}\left(t\right)+\eta^{2}-C_{yy}\left(t\right)\right]-\ln\eta^{2},
\]
which follows from Eqs. (\ref{eq:C11}) and (\ref{eq:C22}).

In the process of performing the trace in Eq. \eqref{eq:Q2}, we define
the second-order functions $C_{ij}^{\textrm{d},\left(u\right)}\left(t,t'\right)=\textrm{tr}_{b}\left[B_{i}\left(t\right)B_{j}\left(t'\right)\left(\rho_{b,\left(u\right)}^{\textrm{d}}-\rho_{b}\right)\right]$.
The complete set of functions $C_{ij}^{\textrm{d},\left(u\right)}\left(t,t'\right)$
can be represented by the independent functions $C_{ij}^{\textrm{d}_{q}}\left(t,t'\right)$
according to
\begin{align}
C_{ji}^{\textrm{d}_{1}}\left(t,t'\right) & \equiv C_{ji}^{\textrm{d},\left(1\right)}\left(t,t'\right)=C_{ij}^{\textrm{d},\left(1\right)}\left(t',t\right)^{*},\label{eq:Cji^d1(t,t')}\\
C_{ji}^{\textrm{d}_{2}}\left(t,t'\right) & \equiv C_{ji}^{\textrm{d},\left(2\right)}\left(t,t'\right)=C_{ij}^{\textrm{d},\left(3\right)}\left(t',t\right)^{*},\label{eq:Cji^d2(t,t')}\\
C_{ji}^{\textrm{d},\left(3\right)}\left(t,t'\right) & =C_{ij}^{\textrm{d},\left(2\right)}\left(t',t\right)^{*}=\left(-1\right)^{\left(\delta_{j1}+\delta_{1i}\right)}C_{ji}^{\textrm{d},\left(2\right)}\left(t,t'\right),\label{eq:Cji^d3(t,t')}\\
C_{ji}^{\textrm{d},\left(4\right)}\left(t,t'\right) & =C_{ij}^{\textrm{d},\left(4\right)}\left(t',t\right)^{*}=\left(-1\right)^{\left(\delta_{j1}+\delta_{1i}\right)}C_{ji}^{\textrm{d},\left(1\right)}\left(t,t'\right).\label{eq:Cji^d4(t,t')}
\end{align}
Writing $\tau=t-t'$, the independent functions $C_{ij}^{\textrm{d}_{q}}\left(t,t'\right)$
are given by
\begin{eqnarray}
C_{11}^{\textrm{d}_{q}}\left(t,t'\right) & = & \eta^{-2}\left[C_{1}^{\textrm{d}_{q}}\left(t\right)+\eta\right]C_{1}^{\textrm{d}_{q}}\left(t'\right)C_{11}\left(\tau\right)\nonumber \\
 &  & +\eta^{-1}C_{1}^{\textrm{d}_{q}}\left(t\right)C_{11}\left(\tau\right)\nonumber \\
 &  & +\eta^{-2}C_{2}^{\textrm{d}_{q}}\left(t\right)C_{2}^{\textrm{d}_{q}}\left(t'\right)C_{22}\left(\tau\right)\nonumber \\
 &  & +C_{1}^{\textrm{d}_{q}}\left(t\right)C_{1}^{\textrm{d}_{q}}\left(t'\right),\label{eq:C_11^dq(t,t')}\\
C_{22}^{\textrm{d}_{q}}\left(t,t'\right) & = & \eta^{-2}\left[C_{1}^{\textrm{d}_{q}}\left(t\right)+\eta\right]C_{1}^{\textrm{d}_{q}}\left(t'\right)C_{22}\left(\tau\right)\nonumber \\
 &  & +\eta^{-1}C_{1}^{\textrm{d}_{q}}\left(t\right)C_{22}\left(\tau\right)\nonumber \\
 &  & +\eta^{-2}C_{2}^{\textrm{d}_{q}}\left(t\right)C_{2}^{\textrm{d}_{q}}\left(t'\right)C_{11}\left(\tau\right)\nonumber \\
 &  & +C_{2}^{\textrm{d}_{q}}\left(t\right)C_{2}^{\textrm{d}_{q}}\left(t'\right),\\
C_{12}^{\textrm{d}_{q}}\left(t,t'\right) & = & \eta^{-2}\left[C_{1}^{\textrm{d}_{q}}\left(t\right)+\eta\right]C_{2}^{\textrm{d}_{q}}\left(t'\right)C_{11}\left(\tau\right)\nonumber \\
 &  & -\eta^{-2}C_{2}^{\textrm{d}_{q}}\left(t\right)C_{1}^{\textrm{d}_{q}}\left(t'\right)C_{22}\left(\tau\right)\nonumber \\
 &  & -\eta^{-1}C_{2}^{\textrm{d}_{q}}\left(t\right)C_{22}\left(\tau\right)\nonumber \\
 &  & +C_{1}^{\textrm{d}_{q}}\left(t\right)C_{2}^{\textrm{d}_{q}}\left(t'\right),
\end{eqnarray}
\begin{align}
C_{31}^{\textrm{d}_{q}}\left(t,t'\right) & =C_{3}^{\textrm{d}_{q}}\left(t\right)C_{1}^{\textrm{d}_{q}}\left(t'\right)-\eta^{-1}C_{2}^{\textrm{d}_{q}}\left(t'\right)C_{32}\left(\tau\right),\\
C_{32}^{\textrm{d}_{q}}\left(t,t'\right) & =C_{3}^{\textrm{d}_{q}}\left(t\right)C_{2}^{\textrm{d}_{q}}\left(t'\right)+\eta^{-1}C_{1}^{\textrm{d}_{q}}\left(t'\right)C_{32}\left(\tau\right),\\
C_{33}^{\textrm{d}_{q}}\left(t,t'\right) & =C_{3}^{\textrm{d}_{q}}\left(t\right)C_{3}^{\textrm{d}_{q}}\left(t'\right).\label{eq:C_33^dq(t,t')}
\end{align}
Thus, all second-order functions $C_{ij}^{\textrm{d},\left(u\right)}\left(t,t'\right)$
appearing in the inhomogeneous terms can be expressed in terms of
$\eta$, $C_{i}^{\textrm{d}_{q}}\left(t\right)$ and $C_{ij}\left(t\right)$.

By inserting $C_{ij}^{\textrm{d},\left(u\right)}\left(t,t'\right)$
from Eqs. (\ref{eq:Cji^d1(t,t')})-(\ref{eq:C_33^dq(t,t')}) into
Eq. \eqref{eq:Dij^d(t,t')}, the complete set of functions $D_{ij}^{\textrm{d},\left(u\right)}\left(t\right)$
can be represented by the two independent sets $D_{ij}^{\textrm{d}_{q}}\left(t\right)$,
according to
\begin{eqnarray}
D_{ij}^{\textrm{d}_{1}}\left(t\right) & \equiv & D_{ij}^{\textrm{d},\left(1\right)}\left(t\right)=\left(-1\right)^{\left(\delta_{j1}+\delta_{1i}\right)}D_{ij}^{\textrm{d},\left(4\right)}\left(t\right),\label{eq:Dji^d1(t,t')}\\
D_{ij}^{\textrm{d}_{2}}\left(t\right) & \equiv & D_{ij}^{\textrm{d},\left(2\right)}\left(t\right)=\left(-1\right)^{\left(\delta_{j1}+\delta_{1i}\right)}D_{ij}^{\textrm{d},\left(3\right)}\left(t\right).\label{eq:Dji^d2(t,t')}
\end{eqnarray}
The explicit expressions for these independent functions are
\begin{eqnarray}
D_{11}^{\textrm{d}_{q}}\left(t\right) & = & \sum_{m}\left\{ \left[\eta^{-1}C_{1}^{\textrm{d}_{q}}\left(t\right)+1\right]D_{1,11,m,1}^{\textrm{d}_{q}}\left(t\right)\right.\nonumber \\
 &  & +\eta^{-1}C_{1}^{\textrm{d}_{q}}\left(t\right)D_{11,m}\left(t\right)\nonumber \\
 &  & \left.+\eta^{-1}C_{2}^{\textrm{d}_{q}}\left(t\right)D_{2,22,m,1}^{\textrm{d}_{q}}\left(t\right)\right\} \nonumber \\
 &  & +C_{1}^{\textrm{d}_{q}}\left(t\right)D_{1}^{\textrm{d}_{q}}\left(t\right),\label{eq:D_11^dq(t)}\\
D_{22}^{\textrm{d}_{q}}\left(t\right) & = & \sum_{m}\left\{ \left[\eta^{-1}C_{1}^{\textrm{d}_{q}}\left(t\right)+1\right]D_{1,22,m,2}^{\textrm{d}_{q}}\left(t\right)\right.\nonumber \\
 &  & +\eta^{-1}C_{1}^{\textrm{d}_{q}}\left(t\right)D_{22,m}\left(t\right)\nonumber \\
 &  & \left.+\eta^{-1}C_{2}^{\textrm{d}_{q}}\left(t\right)D_{2,11,m,2}^{\textrm{d}_{q}}\left(t\right)\right\} \nonumber \\
 &  & +C_{2}^{\textrm{d}_{q}}\left(t\right)D_{2}^{\textrm{d}_{q}}\left(t\right),\\
D_{12}^{\textrm{d}_{q}}\left(t\right) & = & \sum_{m}\left\{ \left[\eta^{-1}C_{1}^{\textrm{d}_{q}}\left(t\right)+1\right]D_{2,11,m,2}^{\textrm{d}_{q}}\left(t\right)\right.\nonumber \\
 &  & \left.-\eta^{-1}C_{2}^{\textrm{d}_{q}}\left(t\right)\left[D_{1,22,m,2}^{\textrm{d}_{q}}\left(t\right)+D_{22,m}\left(t\right)\right]\right\} \nonumber \\
 &  & +C_{1}^{\textrm{d}_{q}}\left(t\right)D_{2}^{\textrm{d}_{q}}\left(t\right),\\
D_{21}^{\textrm{d}_{q}}\left(t\right) & = & \sum_{m}\left\{ \eta^{-1}C_{2}^{\textrm{d}_{q}}\left(t\right)\left[D_{1,11,m,1}^{\textrm{d}_{q}}\left(t\right)+D_{11,m}\left(t\right)\right]\right.\nonumber \\
 &  & \left.-\left[\eta^{-1}C_{1}^{\textrm{d}_{q}}\left(t\right)+1\right]D_{2,22,m,1}^{\textrm{d}_{q}}\left(t\right)\right\} \nonumber \\
 &  & +C_{2}^{\textrm{d}_{q}}\left(t\right)D_{1}^{\textrm{d}_{q}}\left(t\right),
\end{eqnarray}
\begin{align}
D_{31}^{\textrm{d}_{q}}\left(t\right) & =C_{3}^{\textrm{d}_{q}}\left(t\right)D_{1}^{\textrm{d}_{q}}\left(t\right)-\sum_{m}D_{2,32,m,1}^{\textrm{d}_{q}}\left(t\right),\\
D_{32}^{\textrm{d}_{q}}\left(t\right) & =C_{3}^{\textrm{d}_{q}}\left(t\right)D_{2}^{\textrm{d}_{q}}\left(t\right)+\sum_{m}D_{1,32,m,2}^{\textrm{d}_{q}}\left(t\right),\\
D_{13}^{\textrm{d}_{q}}\left(t\right) & =C_{1}^{\textrm{d}_{q}}\left(t\right)D_{3}^{\textrm{d}_{q}}\left(t\right)-\eta^{-1}C_{2}^{\textrm{d}_{q}}\left(t\right)\sum_{m}D_{23,m}\left(t\right),\\
D_{23}^{\textrm{d}_{q}}\left(t\right) & =C_{2}^{\textrm{d}_{q}}\left(t\right)D_{3}^{\textrm{d}_{q}}\left(t\right)+\eta^{-1}C_{1}^{\textrm{d}_{q}}\left(t\right)\sum_{m}D_{23,m}\left(t\right),\\
D_{33}^{\textrm{d}_{q}}\left(t\right) & =C_{3}^{\textrm{d}_{q}}\left(t\right)D_{3}^{\textrm{d}_{q}}\left(t\right).\label{eq:D^dq_33(t)}
\end{align}
As a result, to resolve $D_{ij}^{\textrm{d},\left(u\right)}\left(t\right)$
we express $D_{ij}^{\textrm{d}_{q}}\left(t\right)$ using Eqs. (\ref{eq:D_11^dq(t)})-(\ref{eq:D^dq_33(t)}).
In these expressions, the factors $C_{i}^{\textrm{d}_{q}}\left(t\right)$,
which depend only on $t$, are separated from the functions $C_{i}^{\textrm{d}_{q}}\left(t'\right)$
and $C_{ij}\left(\tau\right)$ that retain the integration variable.
The latter are incorporated into the integral functions $D_{l,ij,m,n}^{\textrm{d}_{q}}\left(t\right)$
and $D_{i}^{\textrm{d}_{q}}\left(t\right)$, as defined in Eqs. (\ref{D^dq_l,ij,m,n(t)})
and (\ref{D^dq_i(t)}), respectively. Consequently, the time-convolution
terms in Eq. \eqref{eq:Dij^d(t,t')} can be evaluated through the
time-local differential equations in Eqs. (\ref{dot_D_ij,m(t)})-(\ref{dot_D_^dq_i(t)}).

\section{Analytical solution of the pure-dephasing model}

\label{sec:appendix-pure_dephasing}

In this section, we show in detail that, in the pure-dephasing limit,
Eq.~(\ref{eq:sigma_P_Q}) reduces to the exact analytical solution
for the pure-dephasing dynamics. In the pure-dephasing limit, where
$\Delta=0$, $\xi_{k}=F(\omega_{k})=1$ {[}see Eq. (\ref{xi_k_DC}){]}
and $H_{sb}'=0$ {[}see Eq. (\ref{H'_sb}){]}, the conditions $\varLambda_{1}=\varLambda_{2}=0$
and $B_{3}=0$ lead to $C_{3}^{\textrm{d},\left(u\right)}\left(t\right)=0$
and $D_{ij}\left(t\right)=D_{ij}^{\textrm{d},\left(u\right)}\left(t\right)=0$
{[}see Eqs.~ (\ref{D_ij(t)})-(\ref{eq:Dij^d(t,t')}){]}. We therefore
write Eq.~(\ref{eq:sigma_P_Q}) explicitly as
\begin{eqnarray}
\left\langle \sigma_{-}\left(t\right)\right\rangle  & = & \textrm{tr}_{s}\left\{ \eta\sigma_{-}\rho_{s}'\left(t\right)\right.\nonumber \\
 &  & +\sum_{u=1}^{4}\left[\sigma_{-}C_{1}^{\textrm{d},\left(u\right)}\left(t\right)\mathcal{G}_{s}'\left(t,0\right)\rho_{s,\left(u\right)}'\right.\nonumber \\
 &  & \left.\left.+i\sigma_{-}C_{2}^{\textrm{d},\left(u\right)}\left(t\right)\mathcal{G}_{s}'\left(t,0\right)\rho_{s,\left(u\right)}'\right]\right\} .\label{eq:sigma_PD}
\end{eqnarray}
Upon performing the trace and using $H_{T}'=H_{s}'+H_{b}$ and $H_{s}'=\frac{\epsilon}{2}\sigma_{z}$,
together with the cyclic property of the trace, the first term in
Eq. (\ref{eq:sigma_PD}) can be evaluated as
\begin{eqnarray}
\textrm{tr}_{s}\left[\eta\sigma_{-}\rho_{s}'\left(t\right)\right] & = & \textrm{tr}_{s}\left\{ \eta\sigma_{-}\textrm{tr}_{b}\left[e^{-iH_{T}'t}\rho_{T}'\left(0\right)e^{iH_{T}'t}\right]\right\} \nonumber \\
 & = & \textrm{tr}_{s}\left\{ \eta e^{iH_{s}'t}\sigma_{-}e^{-iH_{s}'t}\textrm{tr}_{b}\left[e^{-iH_{b}t}\rho_{T}'\left(0\right)e^{iH_{b}t}\right]\right\} \nonumber \\
 & = & \textrm{tr}_{s}\left[\eta\sigma_{-}e^{-i\epsilon t}\rho_{s}'\left(0\right)\right]\nonumber \\
 & = & \eta^{2}\rho_{10}e^{-i\epsilon t}.
\end{eqnarray}
Note that, after performing the trace over the system degrees of freedom,
only the term corresponding to $u=3$ in $\rho_{s}'\left(0\right)$,
as expressed in Eq. (\ref{eq:rhos3_p}), survives. Similarly, for
the remaining terms in Eq. (\ref{eq:sigma_PD}), the system operators
within the trace are given by
\begin{eqnarray}
\textrm{tr}_{s}\left[\sigma_{-}\mathcal{G}_{s}'\left(t,0\right)\rho_{s,\left(u\right)}'\right] & = & \textrm{tr}_{s}\left[\sigma_{-}e^{-iH_{s}'t}\rho_{s,\left(u\right)}'e^{iH_{s}'t}\right]\nonumber \\
 & = & \textrm{tr}_{s}\left[e^{iH_{s}'t}\sigma_{-}e^{-iH_{s}'t}\rho_{s,\left(u\right)}'\right]\nonumber \\
 & = & \textrm{tr}_{s}\left[\sigma_{-}e^{-i\epsilon t}\rho_{s,\left(u\right)}'\right].
\end{eqnarray}
Again, only the term corresponding to $u=3$ survives after performing
the system trace. Subsequently, using Eqs.~ (\ref{eta(w)}), Eq.
(\ref{r(tau)}), (\ref{eq:C_1^d_2(t)}), and (\ref{eq:C_2^d_2(t)}),
Eq. (\ref{eq:sigma_PD}) becomes
\begin{eqnarray}
 &  & \left\langle \sigma_{-}\left(t\right)\right\rangle \nonumber \\
 & = & \left[\eta+C_{1}^{\textrm{d},\left(3\right)}\left(t\right)+iC_{2}^{\textrm{d},\left(3\right)}\left(t\right)\right]\eta\rho_{10}e^{-i\epsilon t}\nonumber \\
 & = & \eta\left[\cosh\left(\textrm{Re}\left\{ r\left(t\right)\right\} \right)-\sinh\left(\textrm{Re}\left\{ r\left(t\right)\right\} \right)\right]\eta\rho_{10}e^{-i\epsilon t}\nonumber \\
 & = & \eta^{2}e^{-\textrm{Re}\left\{ r\left(t\right)\right\} }\rho_{10}e^{-i\epsilon t}\nonumber \\
 & = & e^{-4\int_{0}^{\infty}d\omega\frac{J\left(\omega\right)}{\omega^{2}}\coth\left(\frac{\beta\omega}{2}\right)\left[1-\cos\left(\omega\tau\right)\right]}\rho_{10}e^{-i\epsilon t},
\end{eqnarray}
which recovers Eq.~(\ref{eq:PD_analytical_solution}) in the main
text. References \citep{Breuer02,pure_dephasing_analytical_solution_Palma1996}
have derived the evolution of the off-diagonal matrix element $\tilde{\rho}_{10}\left(t\right)=\left\langle 1\right|\tilde{\rho}_{s}\left(t\right)\left|0\right\rangle $
in the interaction picture, where $\tilde{\rho}_{s}\left(t\right)=e^{\frac{i}{\hbar}H_{s}t}\rho_{s}\left(t\right)e^{-\frac{i}{\hbar}H_{s}t}$.
The expectation value of $\sigma_{-}$ can then be expressed as
\begin{eqnarray}
\left\langle \sigma_{-}\left(t\right)\right\rangle  & = & \textrm{tr}_{s}\left[e^{\frac{i}{\hbar}H_{s}t}\sigma_{-}e^{-\frac{i}{\hbar}H_{s}t}\tilde{\rho}_{s}\left(t\right)\right]\nonumber \\
 & = & \textrm{tr}_{s}\left[\sigma_{-}e^{-i\epsilon t}\tilde{\rho}_{s}\left(t\right)\right]\nonumber \\
 & = & e^{-i\epsilon t}\tilde{\rho}_{10}\left(t\right).
\end{eqnarray}
Thus, by expressing $\left\langle \sigma_{-}\left(t\right)\right\rangle $
in terms of the interaction-picture matrix element $\tilde{\rho}_{10}\left(t\right)$
with the factor $e^{-i\epsilon t}$, Eq. (\ref{eq:PD_analytical_solution})
can likewise be obtained.

\section{Benchmark against HEOM}

\label{sec:appendix_HEOM_benchmark}

\begin{figure}
\includegraphics[width=1\columnwidth]{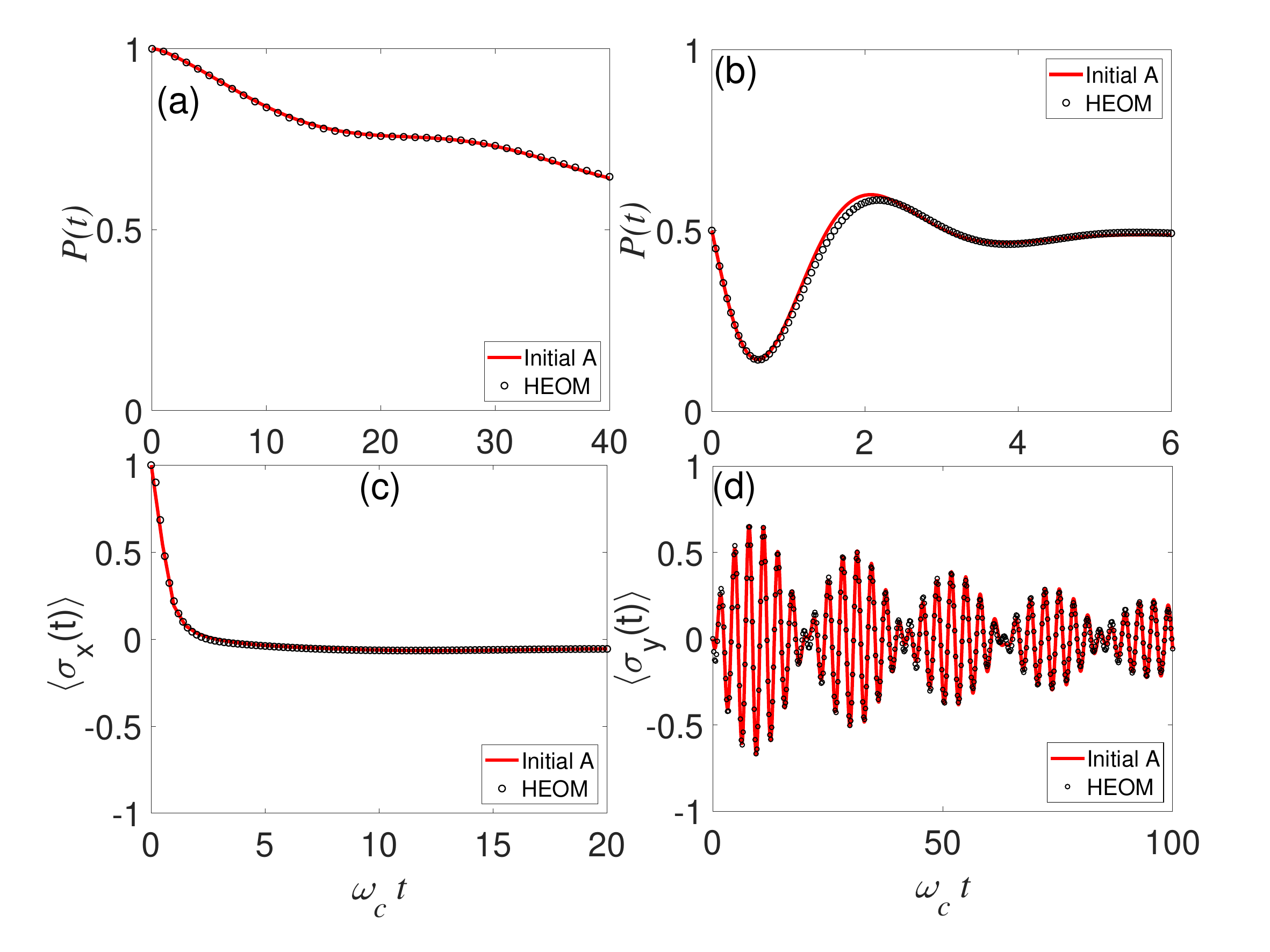}\caption{Comparison between the results obtained from the variational polaron
master equation with the complete transformed initial state (Initial-A;
red solid lines) and the numerically exact hierarchical equations
of motion (HEOM; black open circles). Panels (a)--(d) use the same
initial states and parameters as Figs. \ref{fig:P(t)_spin-boson_T_4}(a),
\ref{fig:P(t)_spin-boson_T_4}(d), \ref{fig:coherence_dynamics}(a),
and \ref{fig:driven_coherence_dynamics}(b), respectively. Panels
(a) and (b) show the population dynamics $P\left(t\right)$, panel
(c) shows $\left\langle \sigma_{x}\left(t\right)\right\rangle $,
and panel (d) shows $\left\langle \sigma_{y}\left(t\right)\right\rangle $
for the driven spin-boson model.}
\label{fig:InitialA_bench_HEOM}
\end{figure}

To illustrate the HEOM benchmark, Fig. \ref{fig:InitialA_bench_HEOM}
presents four representative comparisons between the Initial-A results
and numerically exact HEOM calculations \citep{HEOM_Huang2023,QuantumToolbox.jl2025},
corresponding to the parameter sets used in Figs. \ref{fig:P(t)_spin-boson_T_4}(a),
\ref{fig:P(t)_spin-boson_T_4}(d), \ref{fig:coherence_dynamics}(a),
and \ref{fig:driven_coherence_dynamics}(b). Panels (a) and (b) show
the population dynamics $P\left(t\right)$, panel (c) shows $\left\langle \sigma_{x}\left(t\right)\right\rangle $,
and panel (d) shows $\left\langle \sigma_{y}\left(t\right)\right\rangle $
for the driven spin-boson model. Excellent agreement is observed in
all four cases. The other Initial-A results presented in Secs. V and
VI have also been compared with HEOM calculations and show similarly
good agreement. These comparisons support the reliability of the present
approach in the parameter regimes considered in this work.

\bibliographystyle{apsrev4-2}
\bibliography{RWA_polaron_inhomo_reference}

\end{document}